\documentclass[usenatbib,referee]{mn2e}
\usepackage{amsmath}
\usepackage{graphicx}
\usepackage{epsfig}
\usepackage{txfonts}
\usepackage{color}
\usepackage{harpoon}
\usepackage{multirow}
\usepackage{supertabular}
\usepackage{longtable}

\title[Formation of UMCOWDs from the merger of CO and He WD pairs]
{Formation of ultra-massive carbon-oxygen white dwarfs from the merger of carbon-oxygen and helium white dwarf pairs}
\author[C. Wu, H. Xiong \& X. Wang]
{Chengyuan Wu$^{\rm 1,2,3,4}$\thanks{E-mail:wuchengyuan@ynao.ac.cn}, Heran Xiong$^{\rm 5}$\thanks{E-mail:heran.xiong@anu.edu.au}, Xiaofeng Wang$^{\rm 4,6}$\thanks{E-mail:wang$\_$xf@mail.tsinghua.edu.cn}\\
$^1$Yunnan Observatories, Chinese Academy of Sciences, Kunming 650216, China\\
$^2$Key Laboratory for the Structure and Evolution of Celestial Objects, Yunnan Observatories, CAS, Kunming 650216, China\\
$^3$University of Chinese Academy of Sciences, Beijing 100049, China\\
$^4$Physics Department and Tsinghua Center for Astrophysics (THCA), Tsinghua University, Beijing, 100084, China\\
$^5$Research School of Astronomy and Astrophysics, The Australian National University, Canberra, ACT 2611, Australia\\
$^6$Beijing Planetarium, Beijing Academy of Sciences and Technology, Beijing, 100044, China\\}

\begin{document}
%\date{Accepted. Received}
\date{}
\pagerange{\pageref{firstpage}--\pageref{lastpage}} \pubyear{2021}
\maketitle

\label{firstpage}

\begin{abstract}\label{0. abstract}
Ultra-massive white dwarfs (UMWDs) with masses larger than $1.05{M}_{\odot}$ are basically believed to harbour oxygen-neon (ONe) cores. Recently, Gaia data reveals an enhancement of UMWDs on Hertzsprung-Russell diagram (HRD), which indicates that extra cooling delay mechanism such as crystallization and elemental sedimentation may exist in the UMWDs. Further studies suggested that a portion of UMWDs should have experienced pretty long cooling delays, implying that they are carbon-oxygen (CO) WDs. However, the formation mechanism of these UMCOWDs is still under debate. In this work, we investigated whether the merges of massive CO WDs with helium WDs (He WDs) can evolve to UMCOWDs. By employing stellar evolution code MESA, we construct double WD merger remnants to investigate their final fates. We found that the post-merger evolution of the remnants are similar to R CrB stars. The helium burning of the He shell leads to the mass growing of the CO core at a rate from $2.0\times{10}^{-6}$ to $5.0\times{10}^{-6}\,{M}_\odot\,\rm{yr}^{-1}$. The final CO WD mass is influenced by the wind-mass-loss rate during the post-merger evolution, and cannot exceed about $1.2{M}_\odot$. The remnants with core mass larger than $1.2{M}_\odot$ will experience surface carbon ignition, which may finally end their lives as ONe WDs. Current results implies that at least a portion of UMWDs which experience extra long cooling delay may stem from merging of CO WDs and He WDs.
\end{abstract}

\begin{keywords}
stars: evolution -- binaries: close -- stars: white dwarfs
\end{keywords}

\section{Introduction} \label{1. Introduction}

White dwarfs (WDs) are the most common evolutionary fate of single stars with initial masses lower than $11-12{M}_{\odot}$ (e.g. Nomoto \& Kondo 1991; Siess 2010). Most of these stars are post-AGB remnants with carbon-oxygen (CO) cores. The mass distribution of WDs displays a main peak around $\sim0.59{M}_{\odot}$ and a small peak at the tail of the distribution around $\sim0.82{M}_{\odot}$ (e.g. Kleinman et al. 2013). According to the stellar evolution models, WDs with masses lower than about $0.45{M}_{\odot}$ are He WDs, and those with masses between $0.45-1.05{M}_{\odot}$ are CO WDs. WDs with masses larger than $1.05{M}_{\odot}$ may harbour oxygen-neon (ONe) cores, which are usually named as ultra-massive WDs (UMWDs).

UMWDs are important in studying various physical process such as AGB/SAGB phase, WD cooling theories and asteroseismology (e.g. Doherty et al. 2017; Althaus et al. 2012; Althaus et al. 2021; C\'orsico et al. 2021). Recent Gaia data revealed a number enhancement of massive and ultra-massive WDs within $150\,{\rm pc}$ from the Hertzsprung-Russell HR diagram (HRD). This branch of ultra-massive WDs are named as ``Q branch'' since a large portion of WDs in the region are DQ WDs (e.g. Gaia Collaboration et al. 2018; Gentile Fusillo et al. 2019). The pile-up across the HRD of the Q branch implies that some physical mechanisms should delay the cooling process of the massive WDs. By using the stellar evolution models, Tremblay et al. (2019) inferred that the number enhancement stems from the latent energy which is released by the core crystallization. Further study from Cheng, Cummings \& M\'enard (2019) investigated the dynamical age of the WDs, and found that extra cooling delay mechanisms besides merger delay or crystallization may exist in the UMWDs such as neutron-rich isotopes sedimentation. By using Markov chain Monte Carlo to constrain the extra cooling delay properties and merger fraction, Cheng, Cummings \& M\'enard (2019) suggested that about $6\%$ massive WDs should experience multi-gigayears ($6$-$8\,\rm{Gyr}$) of extra cooling delays.

It is still unclear how does UMWDs could produce pretty long cooling delays lasting for several gigayears. So far, some studies tried to explain the physical mechanism behind the cooling delay. Blouin et al. (2020) presented a new C/O phase diagram which aimed to improve the modeling of the pile-up structure on the HRD. They found that the latent heat release due to the oxygen sedimentation alone cannot explain the observation. Camisassa et al. (2021) calculated the energy released by sedimentation of $^{\rm 22}{\rm Ne}$, and found that an enhancement of $^{\rm 22}{\rm Ne}$ ($6\%$) can significantly prolong the cooling delays of ultra-massive WDs (e.g. Staff et al. 2012). By counting the number fraction of Q branch WDs in different cooling phases, they argued that a large fraction ($50\%$) of UMWDs are derived from double WD mergers. In addition, Bauer et al. (2020) proposed that solid clusters of $^{\rm 22}{\rm Ne}$ may form in liquid CO background plasma, which can sink faster than individual $^{\rm 22}{\rm Ne}$ ions. In that case, the sedimentation of solid $^{\rm 22}{\rm Ne}$ clusters can release more energy, which can slow down the WD cooling significantly. Moreover, Blouin, Daligault \& Saumon (2021) found that the phase separation of $^{\rm 22}{\rm Ne}$ in crystallizing CO WDs can lead to distillation process that efficiently transports $^{\rm 22}{\rm Ne}$ toward the center of WDs, and release significantly amount of gravitational energy which depends on the abundance of $^{\rm 22}{\rm Ne}$.

Up to now, it seems that Q branch WDs are favour to the CO WDs. This is because (1) crystallization occurs earlier in ONe WD than CO WD, which is difficult to produce number enhancement in the narrow region on the color-magnitude diagram (Bauer et al. 2020); (2) $^{\rm 22}{\rm Ne}$ sedimentation can releases more gravitational energy in CO core than in the ONe counterparts (e.g. Camisassa et al. 2021; Bauer et al. 2020); and (3) the gravitational energy released by $^{\rm 22}{\rm Ne}$ phase separation also needs CO WDs (e.g. Blouin, Daligault \& Saumon 2021). However, it is widely believed that single star evolution can hardly produce CO WDs with masses higher than about $1.05{M}_\odot$ since off-centre carbon ignition could be triggered during the SAGB phase, which can subsequently lead to the formation of ONe WDs (e.g. Siess et al. 2007; Doherty et al. 2017). In addition to the internal $^{\rm 22}{\rm Ne}$ abundance, the composition of atmosphere for UMWDs in Q branch also exists discrepancy. Cheng, Cummings \& M\'enard (2019) estimated the fraction of DQ WDs in the cooling delayed population, and found that about half of the cooling delayed WDs are DQ WDs and other half are DA WDs, implying that the cooling delayed UMWDs may have different formation scenarios which need to be further investigated.

The formation scenarios of UMCOWDs remains mystery. Althaus et al. (2021) recently found that single stars with initial masses between $5$-$8{M}_\odot$ have the opportunity to evolve to ultra-massive CO WDs if rotation and different wind-loss rates during TP-AGB phase are considered. under certain circumstances, UMCOWDs with mass range between $1.05$ and $1.35{M}_\odot$ can be produced, which can cover the mass range of Q branch WDs. This formation channel may potentially explain the DA WDs in the Q branch since the UMCOWDs stemmed from single star evolutions usually have H-rich envelops. For the DQ WDs in Q branch, one may expect that they could descend from double CO WD mergers (e.g. Cheng et al. 2019). However, recent post-merger evolutional simulations of double CO WD merger remnants favored other outcomes (e.g. Schwab, Quataert \& Kasen 2016; Wu, Wang \& Liu 2019; Schwab 2021). Schwab (2021) found that the final fates of merger remnants of double CO WDs and CO-HeCO WDs are similar to the single star evolutions, i.e., off-centre carbon ignition can be triggered if the mass of CO core is larger than $1.05{M}_\odot$ and will finally transforms the entire CO core into ONe component, which indicated that UMCOWDs may not originate from double CO WD mergers.

The coalescence of CO WD with He WD have been widely investigated. In the case for which a less massive CO WD merge with another He WD are related to the R Coronae Borealis (R CrB) stars or extreme He stars, and it may evolve to CO WD with mass typically no more than $0.9{M}_\odot$ (e.g. Staff et al. 2012; Zhang et al. 2014; Lauer et al. 2019; Schwab 2019). On the other hand, in the case that the binary system contain a more massive WD and a He WD, rapid mass-transfer of helium onto CO WD may trigger the double-detonation which relates to the type Ia supernova (e.g. Fink et al. 2010; Wang, Justham \& Han 2013; Shen et al. 2018). However, the merger process are strongly affected by the mass ratio of double WDs (e.g. Tutukov \& Yungelson 1979; Jeffery, Karakas \& Saio 2011; Dan et al. 2011; Zhang et al. 2012), and there exists at least some double WD systems, containing an extremely low mass He WD may avoid He-detonation. In that case, the remnant may expand to He-giant-like object during the merger process (e.g. Dan et al. 2014). The evolutionary outcomes of this kind of remnants were less studied, i.e. whether they are related to non-DA WD with CO composition is still uncertain.

In this work, we investigate the remnants resulted from the merger of massive CO WDs with He WDs, with an attempt to study the post-merger evolutionary processes of merging remnants. The article is organized as follows. In Sect.\,2, we provide our basic assumptions and methods used for constructing of the merger remnants. The evolution process of the remnants are shown in Sect.\,3. We discuss our model uncertainties in Sect.\,4 and summarize in Sect.\,5.

\section{Constructing post-merger remnants}\label{Initial models}

\subsection{Criterion of dynamically stable and unstable mass-trensfer}

Double WD mergers play a pivotal role in the formation of many objects, such as, sdO/B stars, R CrB stars, type Ia supernovae and electron capture supernovae (e.g. Greenstein \& Sargent 1974; Wesemael et al. 1982; Webbink 1984; Iben \& Tutukov 1984; Nomoto \& Kondo 1991). Previous works indicated that mass ratio of two WDs and the efficiency of tidal coupling are two vital parameters in affecting the merger process (e.g. Marsh, Nelemans \& Steeghs 2004). if the mass ratio ${\rm q}={\rm {m}_{\rm 2}}/{\rm {m}_{\rm 1}}$ (${\rm {m}_{\rm 1}}$ is the mass of the more massive WD) is less than the critical value ${q}_{\rm crit}$, where
\begin{equation}
{q}_{\rm crit}=\frac{5}{6}+\frac{1}{2}\frac{{\rm d}\,{\rm ln}{R}}{{\rm d}\,{\rm ln}{M}},
\end{equation}
the mass transfer process can be stable, otherwise the increase of radius of the WDs due to the reduction of mass will exceed the increase in the Roche radius caused by the transfer of angular momentum (${R}$ and ${M}$ in equation 1 are the radius and mass of the mass donor WD in double WD system), resulting in dynamically unstable mass transfer process (Tutukov \& Yungelson 1979; Jeffery, Karakas \& Saio 2011). Under strong coupling, i.e. the synchronization time-scale of spin-orbit coupling, ${\tau}_{\rm s}\rightarrow{0}$, the critical mass ratio ${q}_{\rm crit}\approx\frac{2}{3}$ (e.g. Paczy\'nski 1967; Motl et al. 2007). However, if no angular momentum is transferred from accretor to donor, i.e. the synchronization time-scale of spin-orbit coupling, ${\tau}_{\rm s}\rightarrow{\infty}$, ${q}_{\rm crit}$ could decrease to about 0.2 (e.g. Nelemans et al. 2001).

Hydrodynamical simulations proved that mass-transfer process in some low-mass-ratio binaries will be unstable and experience super-Eddington accretion. For example, Dan et al. (2011) provided a series of smoothed particle hydrodynamics calculations, and found that even though the mass ratio of double WDs is very close to the disk stability limit, the mass-transfer phase could be long lived but still unstable. Their results are in agreement with previous theoretical analysis, which means that for the merger of CO WD + He WD pairs, if the mass of CO WD is less then $1.0{M}_\odot$ and the mass of He WD is greater than $0.2{M}_\odot$, the system could experience dynamical unstable mass transfer rather than appear as AM CVn system (e.g. Marsh, Nelemans \& Steeghs 2004). On the other hand, 3D simulations argued that the merger of CO WD + He WD pairs may trigger surface He detonations if the mass of the primary CO WD is massive, which may prevent the remnant from forming massive CO WD. However, the conditions of He detonation, such as density, temperature and geometry of helium torus are relatively complicated (e.g. Guillochon et al. 2010), and systems with massive CO WD and He WD may have opportunity to avoid surface He detonation, resulting in a wider range in forming R CrB stars (e.g. mass of He WDs can enlarge to about $0.37{M}_\odot$; see Dan et al. 2014; Yungelson \& Kuranov 2017). In the present work, we temporarily do not consider the situation of He detonation and only investigate the evolution of the alternative possibility, i.e. the remnants expand into giants after the merger of double WDs.

\subsection{Merger process}

Dan et al. (2014) investigated the merger properties of double WD systems, and found that the structure of the WD remnants can be divided into four regions: (1) cold core; (2) thermally supported envelope; (3) Keplerian disc and (4) tidal tail, where the mass fraction of different regions is mainly determined by the mass ratio of the progenitor. Zhang et al. (2014) established the configuration of CO WD + He WD merger remnants by using 1D stellar evolution code. In their model, they considered that a portion of material from the He WD is accreted by the primary rapidly and a hot corona forms at the onset of merger process, whereas the rest of the material forms a Keplerian disc surrounding the primary WD. The hot corona that piled-up on the surface of the CO WD later expands to a giant size with a radius exceeding about $200{R}_\odot$, and it then gradually swallows the accretion disc. The merger process, that is named as ``composition merger'', can be simulated by an extremely rapid accreting process onto primary WD, and the structures of WD merger remnants that resulted from rapid mass-accretion process of WDs are successfully used in studying the post-merger evolutions of He WD + He WD, CO WD + He WD, and He WD + MS systems, which are related to some objects like hot subdwarfs (sdO/B) and R CrB stars (e.g. Zhang et al. 2012, 2014, 2017; Yu, Zhang \& L$\ddot{u}$ 2021).

\subsection{Initial WD models}

In this work, we are going to employ stellar evolution code \texttt{MESA} (version: 12115; Paxton et al. 2011, 2013, 2015, 2018, 2019) to construct the merger remnants of more massive CO WDs + He WDs based on the methods similar to those adopted by Zhang et al. (2014), and then to investigate the post-merger evolutions of these remnants and their final outcomes.

Based on the criterion mass ratio predicted by previous works, the mass range of merger double WD pairs investigate in the current work have three limitations as follows: (1) the mass of CO WD is less than $1.05{M}_\odot$, i.e. the most massive CO WD formed in single stellar evolution; (2) the mass of He WD is greater than $0.2{M}_\odot$ in order to make sure the merger process is dynamical unstable; (3) the combined mass of double WD is greater than $1.05{M}_\odot$, which can produce remnants with masses large enough to potentially form UMWDs. We use \texttt{MESA} module \texttt{make\_co\_wd} to create initial CO WD and He WD models. Since we concern about the Q branch WDs with mass higher than $1.05{M}_\odot$, the CO WD and He WD masses we considered in the current work are from $0.7$ to $1.0{M}_\odot$ and from $0.2$ to $0.45{M}_\odot$, respectively.

We create non-rotating pre-main-sequence models with masses between $3.5$-$9.0{M}_\odot$ and with different metallicities (solar Z=0.02, subsolar Z=0.001 and metal-rich Z=0.04), then evolve them until their luminosity drop below $0.1{L}_\odot$. During the evolution, OPAL and HELM EOS has been considered (e.g. Rogers \& Nayfonov 2002; Timmes \& Swesty 2000), and we set mixing length coefficient ${\alpha}_{\rm MLT}=1.9179$ which was calibrated from the sun (e.g. Paxton et al. 2011). The OPAL type $2$ opacity is adopted (e.g., Igleslas \& Rooers 1993; Igleslas \& Rooers 1996), which is applicable to extra carbon and oxygen during the He burning. We use Schwarzschild convectional criterion in which composition gradient, semiconvection and thermohaline mixing are not considered. Overshooting parameter is set to be $0.014$, which consistent with that used in SAGB (e.g. Wolf et al. 2013; Denssienkov et al. 2013; Farmer, Fields \& Timmes 2015). Nuclear reaction rates are from JINA REACLIB (e.g. Cyburt et al. 2010), and the nuclear network we considered in the current work contains elements from hydrogen to silicon, including some key isotopes (e.g. $^{\rm 18}{\rm O}$ and $^{\rm 22}{\rm Ne}$). These nuclear reactions contain most of the main reactions during AGB and SAGB phases. The Reimers' and Bl\"ocker' wind mass-loss formula are adopted when the star evolves to the RGB and AGB phases, respectively, and the corresponding mass-loss efficiencies are set to be $\eta=0.5$ (e.g. Reimers 1975; Bl\"ocker 1995). During the late evolutionary phase, thermal pulse process makes the simulations difficult to converge. In order to simplify the calculations, we artificially increase the mass-loss rate in TP-AGB phase to remove all of the hydrogen-rich envelope, and finally obtained CO WD models.

To construct the He WD models, we evolve a series of $\rm 1.8{M}_\odot$ main-sequence stars up the red giant branch until their helium cores increase to the target masses from $0.2$ to $0.45\,{M}_\odot$, then artificially removes their envelopes by a extremely rapid mass-loss rate (i.e. $2.0\times{10}^{-4}\,{\rm M}_\odot\,{\rm yr}^{-1}$, see Xiong et al. 2017; Wu \& Li 2018). Afterwards, the He cores enters WD cooling sequence. As the progenitors of He WDs are less massive stars, they usually have radiation-equilibrium interior cores during the main sequence phase, which result in similar abundances of different elements in various masses of He WDs. Therefore, we use the abundance profile of the H-depletion core to represent the mean elemental abundance of different He WDs. We show the main elemental abundance of our initial WD models in Table\,1.

\begin{table}
\centering
\caption{Elemental abundance of initial WD models}
\begin{tabular}{c|c|c|c|c|c|c|}     % 8 columns
\hline
  Model & CO WD (Z=0.02) & He WD (Z=0.02) & CO WD (Z=0.001) & He WD (Z=0.001) & CO WD (Z=0.04) & He WD (Z=0.04) \\
  \hline
  Mass & 0.7-1.0 &  0.2-0.45  & 0.7-1.0 & 0.2-0.45 & 0.7-1.0 & 0.2-0.45\\
 \hline
 Isotope (\%)    & & & & & &\\
 $^{4}{\rm He}$  & 0.0    & 98.0153  & 0.0    & 99.88      & 0.0      & 96.07\\
 $^{12}{\rm C}$  & 41.99  & 0.0      & 41.62  & 0.0        & 41.0     & 0.02\\
 $^{14}{\rm N}$  & 0.0    & 0.957    & 0.0    & 0.0625     & 0.0      & 2.31\\
 $^{16}{\rm O}$  & 55.5   & 0.407    & 58.25  & 0.04       & 54.0     & 0.35\\
 $^{18}{\rm O}$  & 0.0    & 0.0      & 0.0    & 0.0        & 0.0      & 0.4\\
 $^{20}{\rm Ne}$ & 0.195  & 0.195    & 0.01   & 0.01       & 0.39     & 0.0\\
 $^{22}{\rm Ne}$ & 1.98   & 0.0157   & 0.1    & 0.0015     & 3.8      & 0.03\\
 $^{24}{\rm Mg}$ & 0.008  & 0.08     & 0.004  & 0.004      & 0.16     & 0.16\\
 $^{28}{\rm Si}$ & 0.33   & 0.33     & 0.016  & 0.0016     & 0.65     & 0.66\\
 \hline
\end{tabular}
\end{table}

\subsection{Structures of remnants}

After obtaining the initial CO WD models, we use them to accrete He-rich material with same elemental abundance obtained from He WD models to construct the merger remnants. At the onset of accretion process, the accretion rate should be relative low (${10}^{-7}$-${10}^{-6}\,{M}_\odot\,\rm{yr}^{-1}$) in order to avoid numerical problems since the surface of the CO WDs are very cold. After the WD accumulate a thin He-shell, we increase the mass-accretion rates gradually until it reaches ${10}^{-2}\,{M}_\odot\,\rm{yr}^{-1}$. During the mass-accretion process, we turn off all nuclear reactions in order to speed up the calculations. This operation cannot affect the thermal structure of the remnant and their subsequent evolution since the timescale of the rapid accretion process is extremely short (${\tau}_{\rm acc}\sim$ several tens of years) comparing with He burning timescale of the merger remnants (${\tau}_{\rm nuc}\sim{10}^{5}\,{\rm yr}$). As the material pile-up onto the surface of WD continuously, the temperature on the base of the He layer increases sharply due to the high accretion rate, and the He envelope begin to expand. At the moment when the mass-accretion rate has just reached ${10}^{-2}\,{M}_\odot\,\rm{yr}^{-1}$, the total mass of the hot corona is very small, typically about $0.01{M}_\odot$. Then we use the models of CO WD with hot corona to accrete with the same accretion rate until the total mass reached the required mass. The subsequent accretion process lasts for about tens of years, and the He layer can expand into a giant size with a radius of about $200{R}_\odot$, which can reproduce the ``composition merger'' described in Zhang et al. (2014).

We present the initial structures of different merger remnants in Table\,2 and a representative example in Fig.\,1. Left panels of Fig.\,1 show the temperature and density profiles of $1.25{M}_\odot$ remnants resulted from different initial masses of accreting CO WDs, whereas the right panels show the similar profiles for those which have $0.9{M}_\odot$ initial CO WD but with different masses of envelopes. The rapid mass-accretion processes lead to the formation of hot coronas, which result in the appearances of peak shapes in the temperature-mass coordinate panels. The maximum temperatures of envelopes increase with the initial core mass of the CO WDs, due to that massive WD has higher gravitational acceleration, and the release of gravitational energy can be more efficiently on the massive WD. By contrast, the maximum temperatures on the base of He envelopes are similar if the remnants have the same initial core mass. This is because the WD quickly adjusts its thermal equilibrium state as the mass-accretion process continues, leading to that the maximum temperature of the envelope is mainly determined by core mass. The bottom panels of Fig.\,1 show that the densities at the position of maximum temperature increase with the total masses of the remnants if they have the same initial core mass, since the WD with more accreted materials can provide greater pressure to the bottom of the envelope, leading to the increase of density at bottom of corona. The maximum temperature and density of the remnants in our grid ranges from ${10}^{8.4}$-${10}^{8.6}{\rm K}$ and ${10}^{3.9}$-${10}^{4.1}{\rm g}\,{\rm {cm}^{-3}}$, respectively. Comparing with the SPH simulations, the conditions of the hot coronas are comparable to those in Dan et al. (2014).

\begin{table}
\centering
\caption{Properties of the merger remnants. First two columns represent CO and He WD masses before merger, and the last three columns represent the maximum temperature and density inside the remnants and the star radius of the remnants, respectively.}
\begin{tabular}{c|c|c|c|c|}     % 8 columns
\hline
  ${M}_{\rm CO}/{M}_\odot$ & ${M}_{\rm He}/{M}_\odot$ & ${\rm log}({\rm {T}_{\rm max}}/{\rm K})$ & ${\rm log}({\rho}_{\rm max}/{\rm g}{\rm {cm}^{-3}})$ & ${\rm log}({\rm R}/{\rm {R}_\odot})$\\
  \hline
  0.7 & 0.35 & 8.281 & 3.505 & 2.338\\
  0.7 & 0.40 & 8.283 & 3.532 & 2.339\\
  0.7 & 0.45 & 8.285 & 3.556 & 2.337\\
  \hline
  0.8 & 0.25 & 8.383 & 3.825 & 2.312\\
  0.8 & 0.30 & 8.375 & 3.741 & 2.321\\
  0.8 & 0.35 & 8.379 & 3.818 & 2.321\\
  0.8 & 0.40 & 8.381 & 3.840 & 2.323\\
  0.8 & 0.45 & 8.381 & 3.855 & 2.324\\
  \hline
  0.9 & 0.20 & 8.456 & 3.775 & 2.280\\
  0.9 & 0.25 & 8.455 & 3.836 & 2.330\\
  0.9 & 0.30 & 8.454 & 3.864 & 2.342\\
  0.9 & 0.35 & 8.455 & 3.925 & 2.344\\
  0.9 & 0.40 & 8.455 & 3.946 & 2.343\\
  0.9 & 0.45 & 8.456 & 3.968 & 2.341\\
 \hline
  1.0 & 0.20 & 8.545 & 4.010 & 2.321\\
  1.0 & 0.25 & 8.544 & 4.080 & 2.322\\
  1.0 & 0.30 & 8.545 & 4.139 & 2.322\\
  1.0 & 0.35 & 8.547 & 4.198 & 2.327\\
  1.0 & 0.40 & 8.547 & 4.216 & 2.330\\
  1.0 & 0.45 & 8.550 & 4.270 & 2.333\\
  \hline
\end{tabular}
\end{table}

\begin{figure}
\begin{center}
\epsfig{file=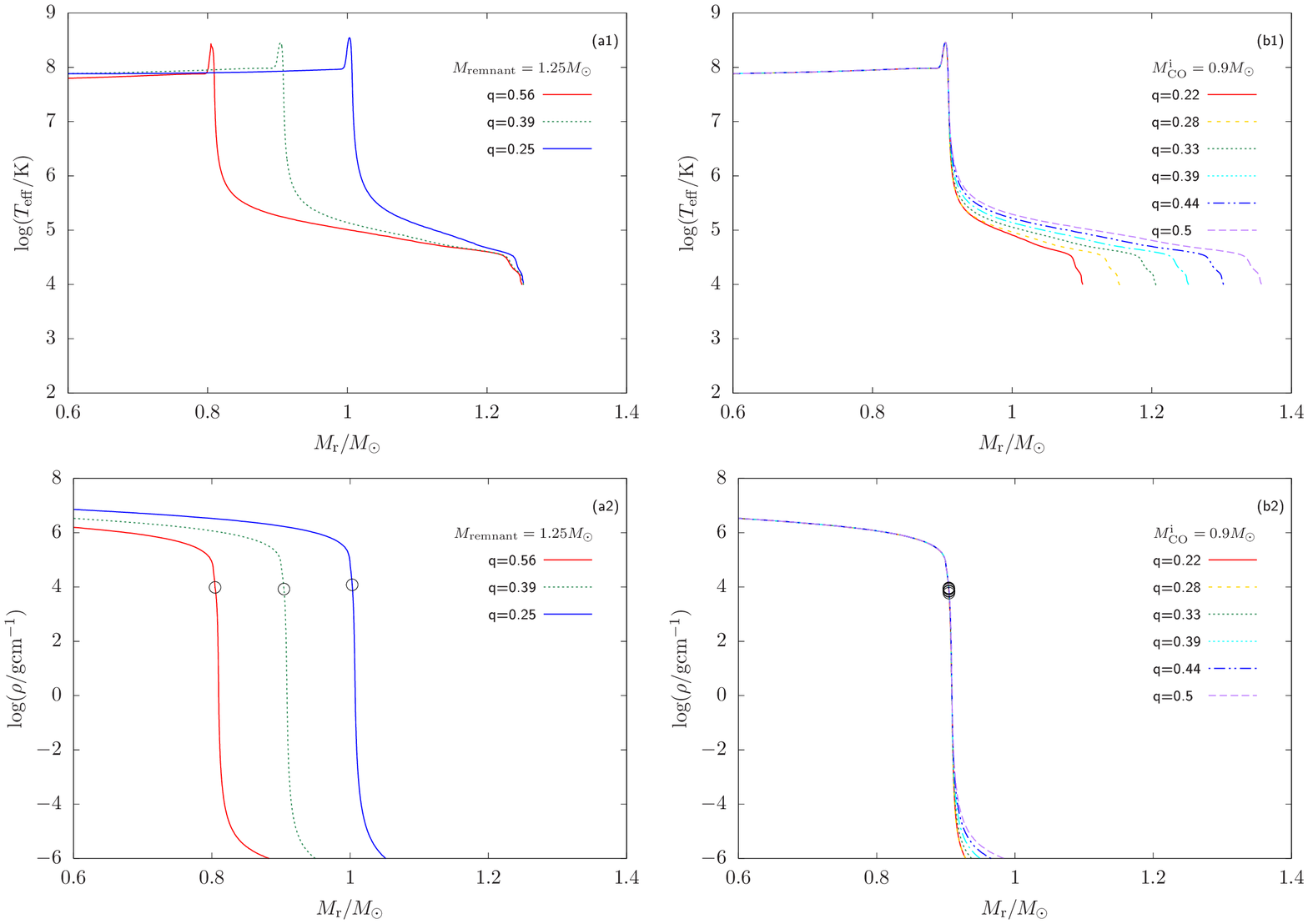,angle=0,width=16.2cm}
 \caption{Temperature/Density-Mass coordinate profiles of different remnants in our calculations. Panels (a1) and (a2) respectively represent the temperature and density profiles of $1.25{M}_\odot$ remnants with different masses of CO cores (${q}={M}_{\rm He}/{M}_{\rm CO}$), whereas panels (b1) and (b2) represent those of the remnants with same CO core mass but different envelope masses. Black circles on the density profiles indicate the corresponding densities at the locations of the highest temperature in the envelopes.}
  \end{center}
\end{figure}

\section{Post-merger evolutions}\label{Evolutions}

Using the method described above, we obtained merger remnants with masses from $1.05$ to $1.45{M}_\odot$ which are resulted from the merger of different initial CO WDs and He WDs. Afterwards, we restart the nuclear reactions and evolve the remnants forward in time.

\subsection{Wind prescription}

The structures of the merger remnants make their consecutive evolutions similar to that of R CrB stars. Wind mass-loss rate is a key parameter in influencing the final fates of the merger remnants since the wind can affect the mass change of envelope and hence the final core mass. However, it is difficult to determine the precise value of wind mass-loss rate because such kind of objects are not well constrained from the observations. Previous works usually artificially limited the wind mass-loss rate of the evolution of merger remnants. For example, Zhang et al. (2014) studied the evolution of R CrB stars that resulted from the merger of less massive CO WD+He WDs. Considering the dusty environment of R CrB stars, they used the Bl\"ocker wind loss prescription with the wind loss factor of $0.02$ (e.g. Bl\"ocker 1995) to calculated the evolution of merger remnants (see also Lauer et al. 2019). They found that the final masses of R CrB stars are about $0.7{M}_\odot$, and the lifetimes of R CrB phases are in the magnitude of about a few ${10}^{4}$ to more than ${10}^{5}\,{\rm yr}$. 

However, the application of Bl\"ocker's wind mass-loss prescription to the R CrB stars may be questioned due to the different compositions in the envelope of R CrB stars comparing with AGB stars. Besides, the terminal velocities of the winds of R CrB stars are in the range of $200$-$350\,{\rm km}\,{\rm {s}^{-1}}$, which is about an order of magnitude higher than that of AGB stars, and therefore extremely restricts the upper limit of mass-loss rate of R CrB stars and their lifetimes, i.e.
\begin{equation}
    \dot{M}_{\rm max}\sim{10}^{-3}\times(\frac{L}{{10}^{4}{L}_\odot})(\frac{{v}_{\infty}}{300\,{\rm km}\,{\rm s}^{-1}})^{-2}\,{M}_\odot\,{\rm yr}^{-1},
\end{equation}
where, $\dot{M}_{\rm max}$, ${\rm L}$ and ${\rm {v}_{\infty}}$ are the upper limit of the mass-loss rate of R CrB stars, luminosity of the stars and the terminal velocities of the winds, respectively (e.g. Clayton, Geballe \&  Bianchi 2003; Schwab 2019). Moreover, it is suggested that R CrB stars are the progenitors of extreme helium (EHe) stars, and the estimated masses of extreme helium stars are higher than the final masses of R CrB stars, which implies that the mass-loss rate in previous models of R CrB stars are overestimated (e.g. Pandey et al. 2001). Schwab (2019) studied the evolutions of R CrB stars by assuming a luminosity and terminal velocity dependent wind mass-loss rate, i.e. $\dot{M}=0.001\times\dot{M}_{\rm max}$, where $\dot{M}_{\rm max}$ is the upper limit of the mass-loss rate of R CrB stars (in this prescription, the mass-loss rate for R CrB stars is in the order of ${10}^{-6}\,{M}_\odot\,{\rm {yr}^{-1}}$; see their equation 1). By comparison, he found that the final mass of R CrB stars is in the range of about $0.6$-$0.7\,{M}_\odot$ by adopting Bl\"ocker's wind mass-loss prescription (${\eta}=0.02$), whereas higher mass can reach if prescription of $\dot{M}=0.001\times\dot{M}_{\rm max}$ is used. Similarly, Schwab (2021) investigated the evolution of CO WD+CO WD mergers, and assumed that the wind mass-loss rate is proportional to the luminosity and radius of the remnant but inversely proportional to the mass, which is similar to the mass-loss rate given in Reimers (1975) since the mass-loss rates of giants with carbon enhanced envelopes are uncertain as well. Therefore, in the current work, we are going to analyze the evolutions of the merger remnants by assuming Reimers' wind mass-loss prescription, and then discuss the uncertainties of mass-loss rates in Sect.\,4.

\subsection{Results of remnants evolutions}

The complete results of the grid in our calculations are listed in Table.\,A1-A3 in Appendix A. In the following examples, we present the evolutionary properties of some representative remnants under the Reimers' wind mass-loss prescription.

Fig.\,2 shows the evolutionary tracks of different remnants on the Hertzsprung-Russell diagram (HRD). The He flashes at the beginning of the evolution occur on the base of the He envelope, resulting in the fluctuations on the HRD. Owing to the initially compact structure of the merger remnant, a small amount of $^{\rm 20}{\rm Ne}$ and $^{\rm 24}{\rm Mg}$ are produced during the He shell flashes phase through the following nuclear reactions:
\begin{equation}
    ^{\rm 12}{\rm C} + ^{\rm 12}{\rm C}\,\rightarrow\,^{\rm 20}{\rm Ne} + ^{\rm 4}{\rm He},
\end{equation}
\begin{equation}
    ^{\rm 12}{\rm C} + ^{\rm 16}{\rm O}\,\rightarrow\,^{\rm 24}{\rm Mg} + ^{\rm 4}{\rm He},
\end{equation}
and
\begin{equation}
    ^{\rm 20}{\rm Ne} + ^{\rm 4}{\rm He}\,\rightarrow\,^{\rm 24}{\rm Mg}.
\end{equation}
However, comparing to the double CO WD merger remnants that have more compact structures, the bottom of the He shells on the CO+He WD merger remnants have relatively lower temperatures and densities, resulting in that the He flashes have not triggered the inwardly propagating carbon flames. After about a hundred of years, the remnant restored thermal equilibrium, and the evolution of remnant is similar to a Helium giant. The continuous He burning transforms He into carbon and oxygen, leading to increase of the degenerate core mass, and the temperature and luminosity of the remnant. For the remnants with more massive initial CO core, the luminosity are higher after the remnants restored the thermal equilibrium, as the evolution of this kind of giant-like objects comply with the core mass-luminosity relation (e.g. Jeffery 1988). For those with similar initial core mass, as shown in panel (b) of Fig.\,2, remnants with more massive He envelopes can evolve to higher luminosity as more CO material can be generated during the He burning phase.

Most of the remnants with initial total masses no more than $1.25{M}_\odot$ can lose all of the He-rich envelope during the giant phase and finally form WDs and evolve along the cooling tracks. However, for those with higher total masses, e.g. ${\rm q}=0.25$ in panel (a) and ${\rm q}=0.5$ in panel (b) of Fig.\,2, the evolution of these remnants show a suddenly increase in luminosity on the HRD. This is because the bottom of the convectional region of the He shell can penetrate into the He burning layer as the increase of the degenerate core, leading to the increase of He abundance in the flame. Owing to that the nuclear reaction rate of He burning is proportional to the square of He abundance (e.g. Denissenkov et al. 2013), the resulted luminosity increases drastically with the energy generation rate. In the case of $q=0.5$ in panel (b) of Fig.\,2, the temperature of the CO surface increases gradually as the CO ashes pile up on the degenerate core, which causes the ignition of surface carbon during its evolution. In this situation, carbon ignition occurs when the core mass increases to about $1.2{M}_\odot$. Due to the great energy release rate of carbon burning, our simulations come across some numerical problem, and we did not follow the subsequent evolution. For cases experiencing carbon ignition, we just deduce that the inwardly propagating carbon flames can transform carbon into neon until they reach the centre or quench somewhere inside the CO core which depend on the convectional boundary mixing (e.g. Schwab et al. 2016; Denissenkov et al. 2013; Wu et al. 2020).

\begin{figure}
\begin{center}
\epsfig{file=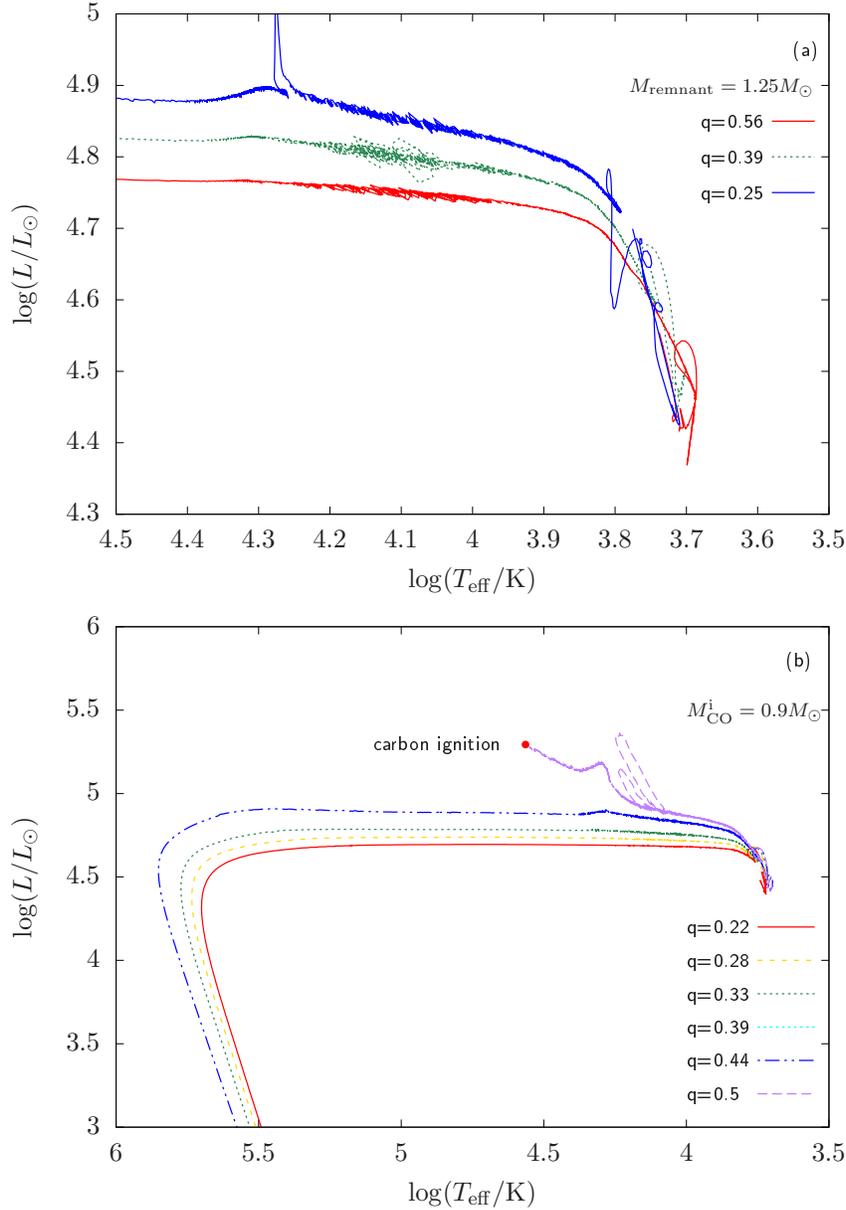,angle=0,width=12.2cm}
 \caption{HR diagram of different remnants. Panel (a): evolutionary tracks of $1.25{M}_\odot$ remnants with different CO core masses on HRD. Panel (b): similar to panel (a), but for different masses of remnants with $0.9{M}_\odot$ CO core. Red point indicates the position where off-centre carbon ignition occurs.}
  \end{center}
\end{figure}

For the remnant with higher total mass, the bottom of the convectional region can extend inward gradually until it reaches the He burning layer when the core mass increase to about $1.15{M}_\odot$. In order to show the variety of the convectional region and how the He burning affect the surface elemental abundance, we present the Kippenhahn diagrams and the mean abundance of different elements in the surface convectional region for three representative examples in Figure 3. Due to the high temperature of the He burning layer, the bottom of the convectional region is very close to the burning layer on the mass coordinate, and the outside of the convectional region can extend to the surface of the star. Owing to the high initial luminosity, the base of the convective region is closer to the burning layer in the case with $0.9{M}_\odot$ CO core than that with $0.8{M}_\odot$ at the beginning of the evolution. This can be seen from the small window in each left panel of Fig.\,3, which shows the temporal change in mass between bottom of the convectional region and the He burning layer (${M}_{\rm conv}-{M}_{\rm esp}$). In the cases of ${\rm q}=0.56$ and $0.39$, the core mass has not reached $1.15{M}_\odot$ before the remnants evolve to WDs, and the inwardly extended convectional region has not reached burning layer. However, for more massive remnant, the convectional region which can affect the surface elemental abundance significantly, as it has reached the burning layer during the evolution.

As shown in right panels of Fig.\,3, the mean elemental abundance of the convectional region will show little change during the evolution if the bottom of the convectional region does not touch the burning layer. But in the cases when the extended convective region penetrated into the burning flame, the helium and nitrogen in the envelope can be transported into the burning layer, which may cause the exhaustion of $^{\rm 14}{\rm N}$ and $^{\rm 18}{\rm O}$ in the convectional envelope through the reactions of $^{\rm 14}{\rm N}$(${\alpha}$,${\gamma}$)$^{\rm 18}{\rm O}$(${\alpha}$,${\gamma}$)$^{\rm 22}{\rm Ne}$. The depletion of $^{\rm 14}{\rm N}$ and $^{\rm 18}{\rm O}$ release considerable amount of energy successively, leading to the rapid increase in luminosity, leading to the formation of two ``loops'' on the HRD (see Fig.\,2). After the $^{\rm 14}{\rm N}$ and $^{\rm 18}{\rm O}$ have been exhausted, the continuous He burning gradually decrease the mean He abundance in the envelope. In this situation, elements in the envelope is dominated by carbon before the surface of CO core ignition occurs.

\begin{figure}
\begin{center}
\epsfig{file=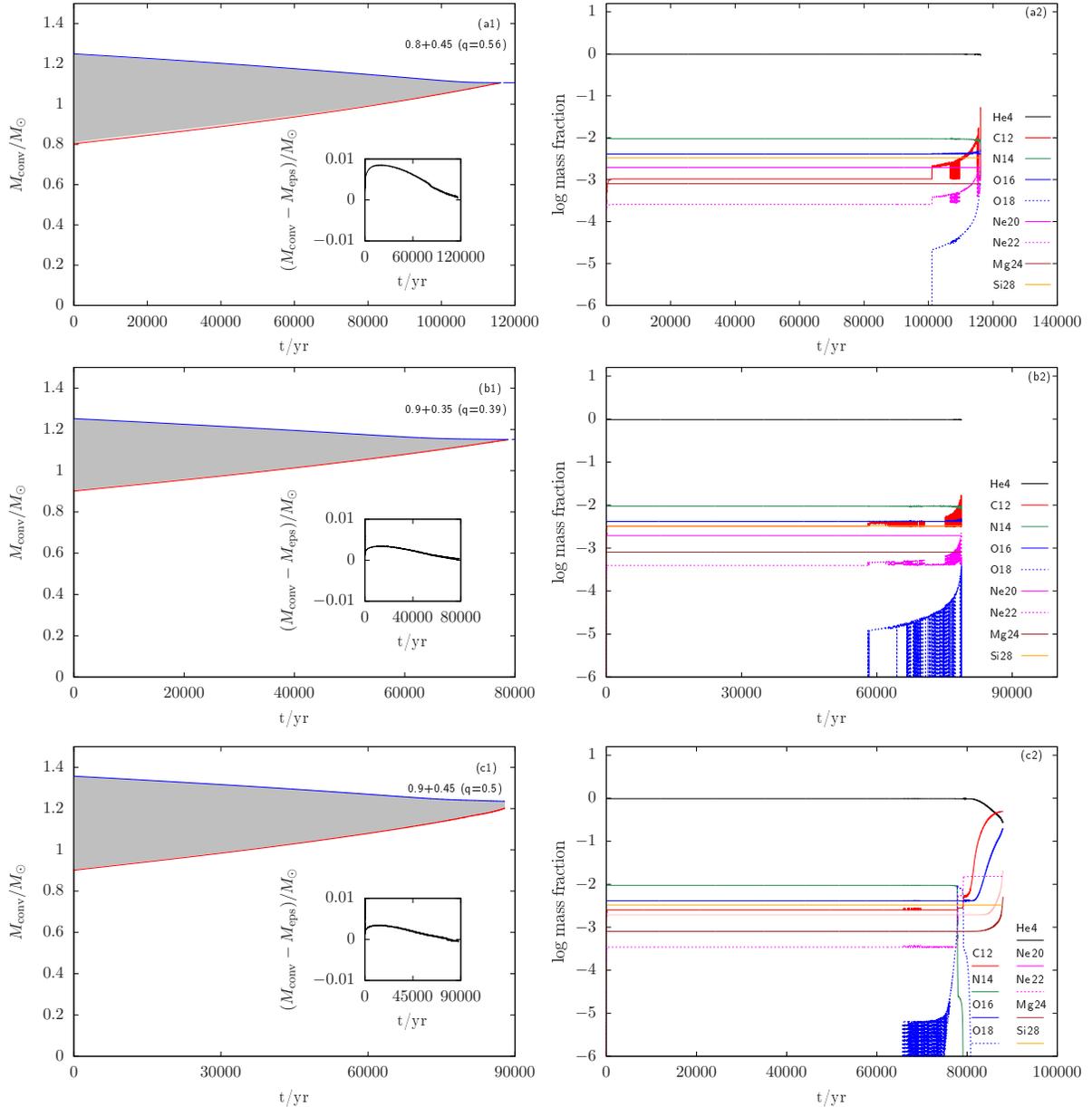,angle=0,width=16.2cm}
 \caption{Kippenhahn diagrams (left panels) and the changes in mean elemental abundances of envelopes over time (right panels) for three different remnants. Grey shadows in Kippenhahn diagrams represent the evolutions of convective regions, whereas blue and red lines represent the changes in total masses and mass coordinates of He flames over time, respectively.}
  \end{center}
\end{figure}

The evolution timescale of remnants depend on both the wind mass-loss rates and the increase rate of core mass. In Fig.\,4, we show the changes in wind mass-loss rate, radius and luminosity of the remnants over time. After about a hundred years of thermal adjustment, the remnant enters the state of thermal equilibrium. The subsequence evolution of the remnant is more like a He-giant, for which the stellar luminosity and radius remains constant for most of its left time. The luminosity in the equilibrium state increases with the core mass according to the core mass-luminosity relation, which can reach the higher luminosity at the final stage of evolution for more massive remnant, whereas the discrepancy of radius among various models is not significant. The mass-loss rate for various remnants is typically on an order of magnitude of ${10}^{-6}\,{M}_\odot\,{\rm {yr}^{-1}}$, which increases with the core mass.

\begin{figure}
\begin{center}
\epsfig{file=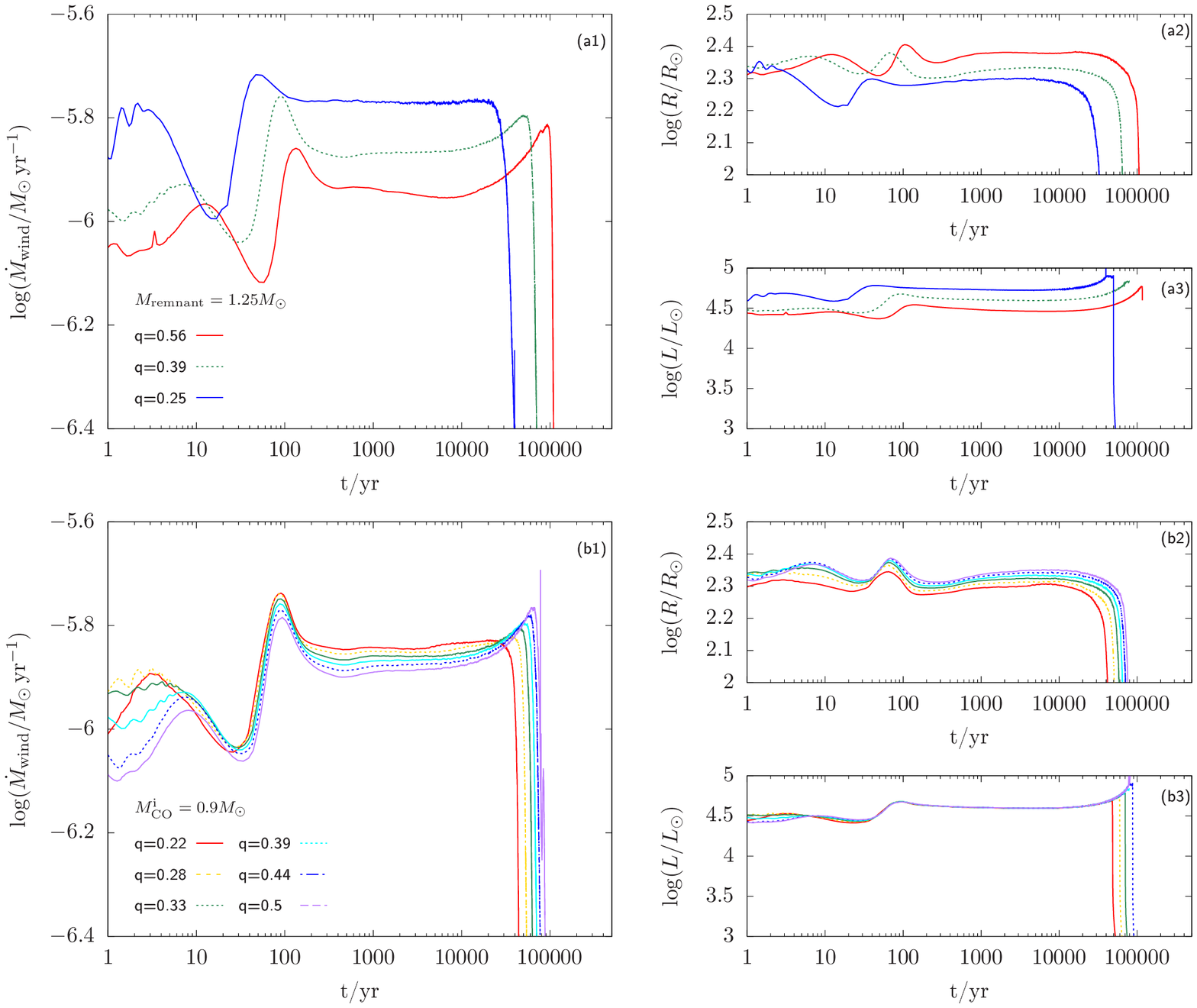,angle=0,width=16.2cm}
 \caption{Properties of some key physical parameters evolve with time. Upper panels: changes in mass-loss rate, radius and luminosity of $1.25{M}_\odot$ remnants with different masses of envelope over time. Lower panels: similar to upper panels, but for those of different masses of remnants with $0.9{M}_\odot$ CO core.}
  \end{center}
\end{figure}

In Fig.\,5, we present the evolution of mass increasing rate of the CO core for different models. The increasing rate of CO core is determined by the core mass because He burning is faster in the remnants with massive core compared with the lower-mass counterparts. For the remnant with more massive envelope, the giant needs more time to exhausts the He envelope, which results in longer evolutionary timescale. The mass increase of the CO core for various mass of remnants lies between $2\times{10}^{-6}$ and $5\times{10}^{-6}\,{M}_\odot\,{\rm {yr}^{-1}}$, which is similar to the conditions in steadily He burning region of He-accretion CO WD (e.g. Wang et al. 2015; Wu \& Wang 2019; Wu et al. 2020).

\begin{figure}
\begin{center}
\epsfig{file=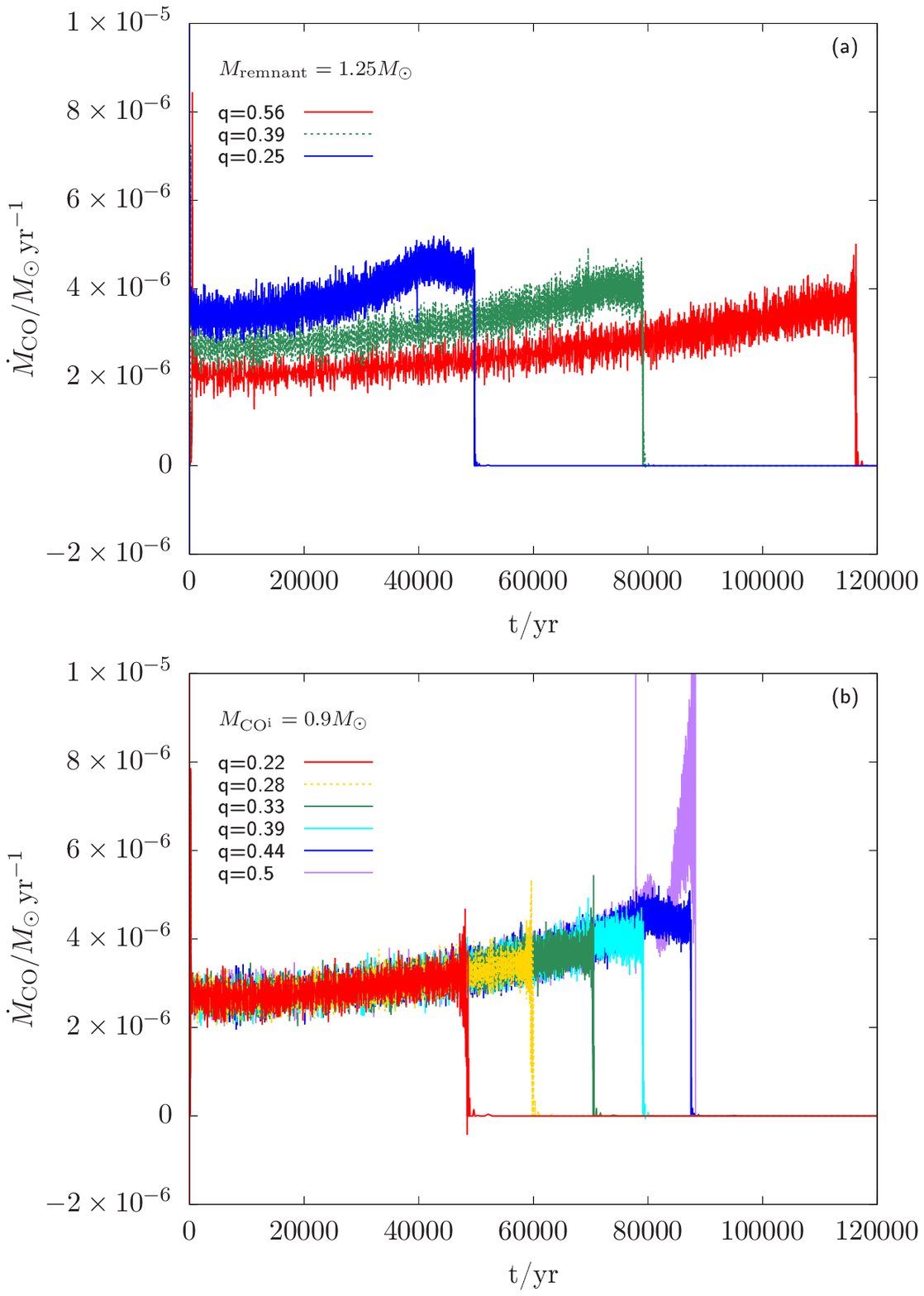,angle=0,width=12.2cm}
 \caption{Changes in mass-accumulation rates of different remnants over time. Panel (a): $1.25{M}_\odot$ remnants with different initial CO core masses. Panel (b): similar to panel (a), but for those of different mass of remnants with $0.9{M}_\odot$ CO core.}
  \end{center}
\end{figure}

As the CO core increases in mass, the surface becomes hotter because of the thermal contract of the core, which may lead to the off-central carbon ignition. In the mass-accretion WD systems, previous works found that if a massive CO WD accretes He-rich material at a rate higher than about $2.5\times{10}^{-6}$ to ${10}^{-5}\,{M}_\odot\,{\rm {yr}^{-1}}$, an off-centre carbon ignition may occur before the WD mass increase to ${M}_{\rm Ch}$ limit (e.g. Kawai, Saio \& Nomoto 1987; Saio \& Nomoto 1998; Brooks et al. 2016; Wang, Podsiadlowski \& Han 2017; Wu \& Wang 2019). The ignited mass of CO core is mainly determined by the mass-accretion rate (inversely proportional to the mass-accretion rate) rather than the initial core mass. In the CO WD + He WD merger systems, the mass increase of the CO core is caused by He burning on the base of the thick envelope; and the He burning rate is mainly determined by the CO core mass. Since the luminosity of various remnants is similar during their evolution, the final core mass trigging off-centre carbon ignition does not have differences for various WD merger. In Fig.\,6, we present the final masses of the remnants at the end of evolutions under different wind mass-loss prescriptions. We found that if the core increases in mass to about $1.2{M}_\odot$, carbon under the helium shell can be ignited, which could subsequently form inwardly propagating carbon flame, leading to the formation of non-CO UMWDs. In this situation, we just recorded the remnant mass when carbon ignition occurs (labelled by filled circles in Fig.\,6), and did not explore the further evolution as the calculations are extremely time consuming. Our result indicates that CO WD+He WD merger can lead to the formation of UMCOWDs but with mass no more than $1.2{M}_\odot$.

\begin{figure}
\begin{center}
\epsfig{file=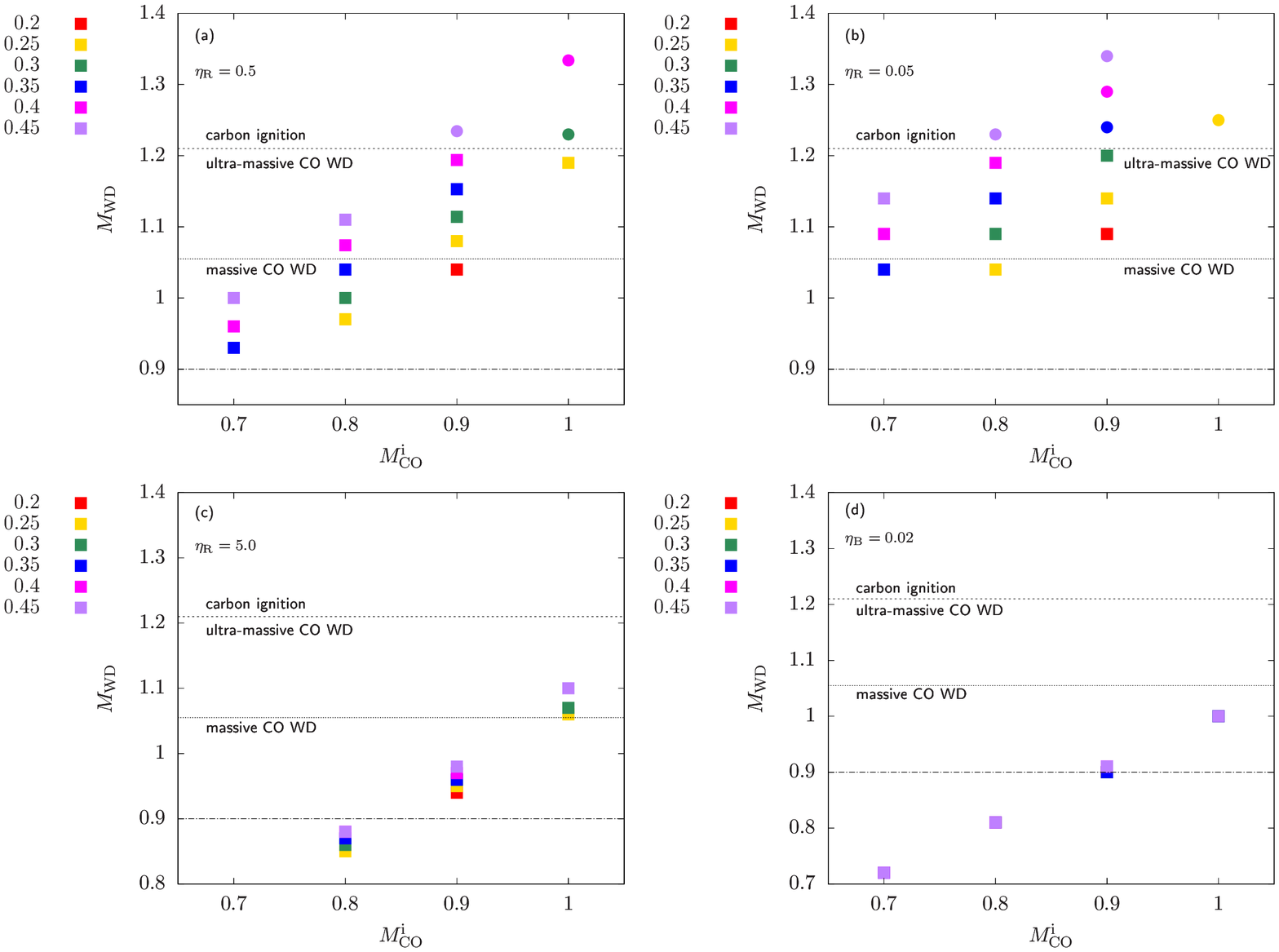,angle=0,width=15.2cm}
 \caption{Initial core mass-final WD mass relation. X-axis represents the initial CO core mass in various remnants, whereas labels with different colours represent the mass of He-rich envelope for different remnants. Labels of square above black dot-dashed line represent the remnants that can eventually evolve to massive CO WDs, whereas circles above black dotted line represent those which experience off-centre carbon ignitions. Panels (a)-(c): under Reimers' wind prescription with various of wind scaling factors. Panel (d): under Bl\"ocker wind prescription.}
  \end{center}
\end{figure}

\section{Discussion}

\subsection{Model uncertainties}

In the previous section, we explored the evolutional fates of the merger remnants under the basic assumptions. However, some uncertainties such as wind mass-loss rate, rotation and initial metallicity of the progenitors may influence the evolution and the final fates of the remnants which needs to be further investigated.

\subsubsection{Wind mass-loss rate}

As analyzed in previous section, wind mass-loss rate is one of the most uncertain parameters in simulating post-merger evolutions. We have shown the evolutionary results by adopting the Reimers' wind formula with the default wind factor of ${\eta}_{\rm R}=0.5$, which predicts the similar wind mass-loss rate in Schwab (2021) of about ${10}^{-6}\,{M}_\odot\,{\rm {yr}^{-1}}$ and similar lifetimes of R CrB phase estimated in Zhang et al. (2014) and Lauer et al. (2019) of about ${10}^{4.7}$ to ${10}^{5.2}\,{\rm yr}$.

In order to investigate the effects of wind mass-loss rate, we utilized different mass-loss rate by either artificially altering the values of ${\eta}_{\rm R}$ or by changing the wind prescription to Bl\"ocker's wind with ${\eta}_{\rm B}=0.02$ during the giant phase, and then evolve the remnants forward in time. In Fig.\,7, we present the evolutionary properties of the $1.25{M}_\odot$ remnant in different wind mass-loss rates. The evolution of the core mass keeps the same in different cases since the He burning rate is determined by the core mass, whereas the evolutionary timescales of the remnants are significantly different due to the influence of mass-loss rate. In the cases of enhanced mass-loss rates, a higher fraction of material is blown away by the wind, resulting in the formation of typically less massive CO WD. In contrast, for the remnant with lower mass-loss rate, a more massive He envelope can be remained at the time when the core increases its mass to $1.15{M}_\odot$, leading to the rapid exhaustion of $^{\rm 14}{\rm N}$ and $^{\rm 18}{\rm O}$ in the envelope, which cause the loops in the HRD. However, our simulations indicate that the carbon ignition mass of the core will not change under different wind assumptions, which means that the maximum mass of UMCOWDs resulted from CO WD+He WD mergers are still $1.2{M}_\odot$.

It is not unanticipated results that the final mass range of CO WD descended from merger of CO WD + He WD pairs and the lifetimes of the merger remnants appear as R CrB stars are significantly affected by different wind mass-loss rates. According to the evolutionary results listed in appendix, the discrepancies of wind mass-loss rates are within an order of magnitude if use different Reimers' wind factors. However, due to the much more steep proportional relationship between star mass and luminosity in Bl\"ocker's wind prescription (i.e. $\dot{{\rm M}}\propto{\rm {L}^{3.7}}$), mass-loss rates of the remnants are extremely higher under the later assumption even though a relatively smaller wind scaling factor are adopted (i.e. ${\eta}_{\rm B}=0.02$). The average wind mass-loss rate can as large as a few of ${10}^{-4}\,{M}_\odot\,{\rm {yr}^{-1}}$, leading to pretty short lifetime of the remnants appear as R CrB stars (the remnants can lost their envelopes within $500$ to $5000\,{\rm yr}$). As a result, the UMCOWDs can be produced through double WD mergers under Reimers' wind prescription even if with large wind scaling factor, but can hardly produced through Bl\"ocker's wind prescription. As analyzed by previous works that H-depleted UMCOWDs should exist in nature, which may imply an indirect limitation on wind mass-loss rate of the R CrB stars, otherwise the observed class of WDs may difficult to be produced.

\begin{figure}
\begin{center}
\epsfig{file=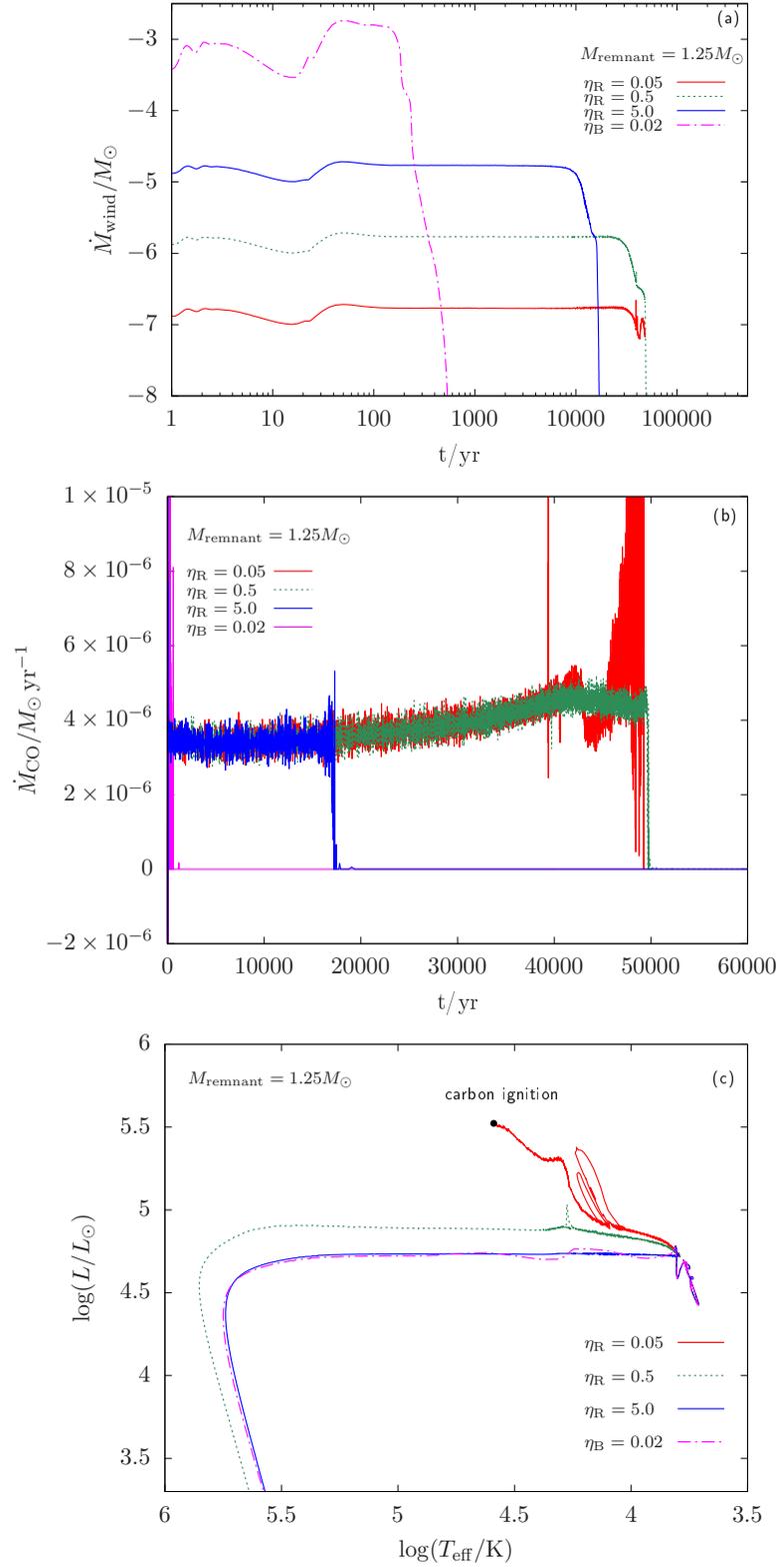,angle=0,width=11.2cm}
 \caption{Panel (a): changes in wind mass-loss rates over time for $1.25{M}_{\odot}$ remnant with $0.9{M}_\odot$ CO core. Three lines represent the evolutions under different wind parameters. Panel (b): similar to panel (a), but for the changes in accumulation rates of CO cores over time. Panel (c): evolutions of the remnants on the HRD. Black point indicates the position where off-centre carbon ignition occurs.}
  \end{center}
\end{figure}

\subsubsection{Rotation}

In the basic assumptions, we did not consider the effect of rotation. However, the merged remnant can remain portion of the orbital angular momentum of the double WD system during the merger process. The rotation may delay the occurrence of off-centre carbon ignition of the CO core due to the lifting effect and influence the wind-loss process during the giant phase (e.g. Yoon, Podsiadlowski \& Rosswog 2007; Schwab 2021). As the envelope of merger remnant expands in radius, it will spin down due to the angular momentum redistribution, which results in differential rotation between the core and envelope. The fast rotating CO core will transform angular momentum into the envelope gradually due to the viscosity effect in the subsequent evolution, which may cause changes in temperature, pressure and convective state of the envelope. In order to investigate the rotational effect, we added rotating torque in different shell of remnant after the He burning become stable until the angular velocity of different shells reached the desired value. In these simulations, we adopt Tayler-Spruit dynamo in MESA default to treat the angular momentum transport (e.g. Paxton et al. 2013), and simply considered that the remnant contains a slowly co-rotating envelope and a faster co-rotating inner core, in which the angular velocity (${\rm \Omega}$) of the envelope and core are $40\%$ of their Keplerian velocity (i.e. ${\rm \Omega}_{\rm core}=0.4\times{\rm \Omega}_{\rm core,crit}$, ${\rm \Omega}_{\rm envelope}=0.4\times{\rm \Omega}_{\rm envelope,crit}$, where ${\rm \Omega}_{\rm crit}=\sqrt{\frac{GM}{{R}^{3}}}$). After configuring the structures of rotating remnants, we remove the rotating torque and then evolve the models forward in time.

In Fig.\,8, we present the HRD of the $1.3{M}_\odot$ remnant resulted from the merger of $1.0{M}_\odot$ CO WD with $0.3{M}_\odot$ He WD, and the corresponding evolution of convective region in the He envelope. The non-rotation case is over plotted for comparison (see blue lines). Owing to the viscosity effect from the differential rotation, the angular momentum transforms from the core to the envelope, which makes the core spin down. The viscosity between the core and envelope is extremely efficient which results in the angular momentum transport process finished in about $110{\rm yr}$. As the angular momentum transforms from core to the envelope, the rotation of the surface of the remnant spins up, leading to a slightly increase of the mass-loss rate of the remnant. This causes change in total mass of the remnant over time, which can be seen from the solid lines in panel (b) of Fig.\,8. In the subsequent evolution, the inner core keeps spinning at a low angular velocity, which causes an effect similar to that from an increase in mass-loss rate. At the end of evolution, the core mass has increased to about $1.2{M}_\odot$, when carbon ignition still occurs due to contraction of the core. Current results indicate that rotation can lead to enhancement of wind mass-loss, but it can hardly alter the final fate of the core.

\begin{figure}
\begin{center}
\epsfig{file=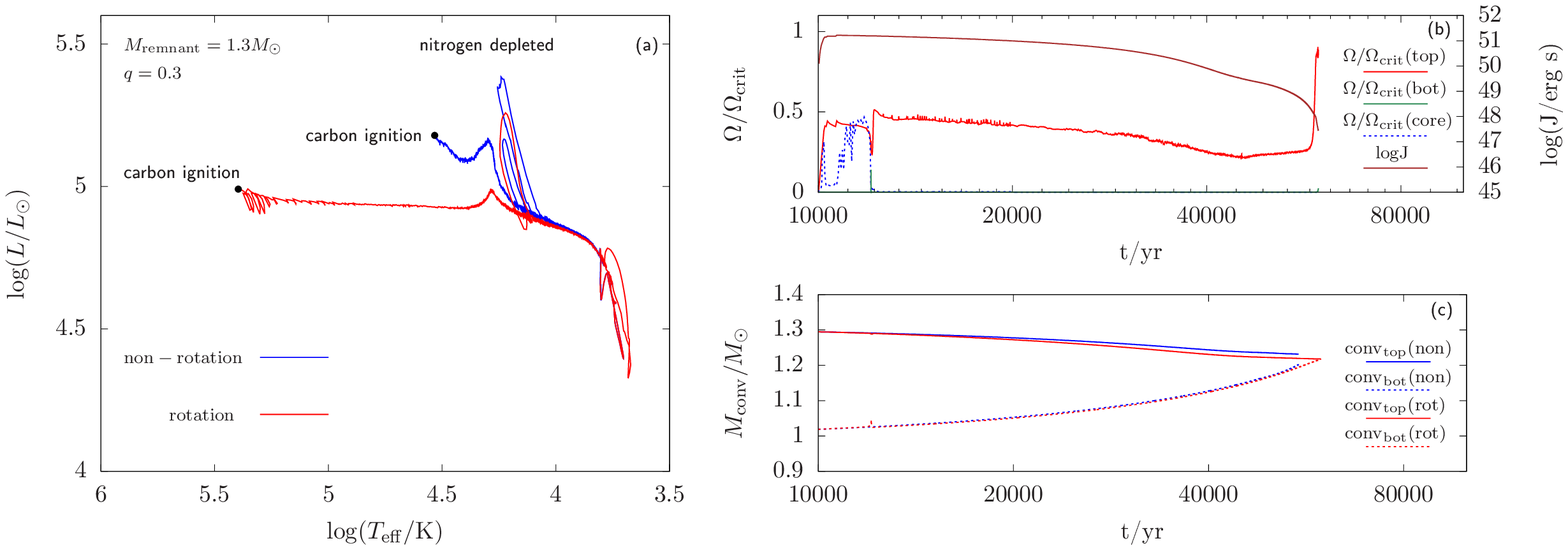,angle=0,width=17.2cm}
 \caption{Properties of the $1.3{M}_\odot$ remnant change over time. Panel (a): brown line represents the evolution of total angular momentum of the remnant, whereas red, green and blue lines represent the changes in angular velocities of surface, bottom of the envelope and surface of the CO core over time, respectively. Panel (b): convective boundaries of the envelope in the cases with (red) and without (blue) rotation. Panel (c): HRD for the models with (red) and without (blue) rotation. Black points indicate the positions where off-centre carbon ignition occur.}
  \end{center}
\end{figure}

\subsubsection{Initial metallicity}

According to cooling models of WD, the UMWDs in Q branch that experience extra cooling delay may stem from CO WDs with an enhancement of $^{\rm 22}{\rm Ne}$ (e.g. Bauer et al. 2020; Camisassa et al. 2021). The enhancement of $^{\rm 22}{\rm Ne}$ in a WD is likely due to that the progenitor of the WD is metal-rich (e.g. Garc\'ia-Berro et al. 2010). Besides, double WD merger process such as the formation of R CrB star may also enhance the abundance of $^{\rm 22}{\rm Ne}$ in the primary WD (e.g. Staff et al. 2012). In order to study the elemental abundance of UMCOWD formed from CO WD + He WD merger in different environments and whether metallicity could affect the evolution of the remnant, we create CO and He WD models with two different metallicities, i.e. Z=0.001 and 0.04. Then construct the merger remnants by using the same methods described in Sect.\,2, and further investigate their evolutions.

We found that there is almost no difference for the evolutional properties of the remnants with different metallicities, such as luminosity, mass-loss rate and He-burning rate. Off-centre carbon ignition will occur when the mass of CO core increases to $1.2{M}_\odot$, which indicates that the initial metallicity of the progenitor cannot change the final fate of the remnant. However, the abundances of $^{\rm 22}{\rm Ne}$ are much different in the WDs for different initial metallicities. In Fig.\, 9, we present the abundance profile of the UMCOWD evolved from $1.25{M}_\odot$ remnants with different metallicities. One can see that the abundance of $^{\rm 22}{\rm Ne}$ increases with the initial metallicities, which is consistent with the trend predicted by single star evolution. The He WDs that produced in metal-rich environment will retain higher abundance of $^{\rm 14}{\rm N}$, which accounts for abundance difference in the envelope of merger remnants. Due to the high temperature of the He burning flame, $^{\rm 14}{\rm N}$ can be completely exhausted and transformed into $^{\rm 22}{\rm Ne}$ during the evolution, which means that the progenitor environment is one of the most important conditions affecting the cooling delay of UMWDs.

\begin{figure}
\begin{center}
\epsfig{file=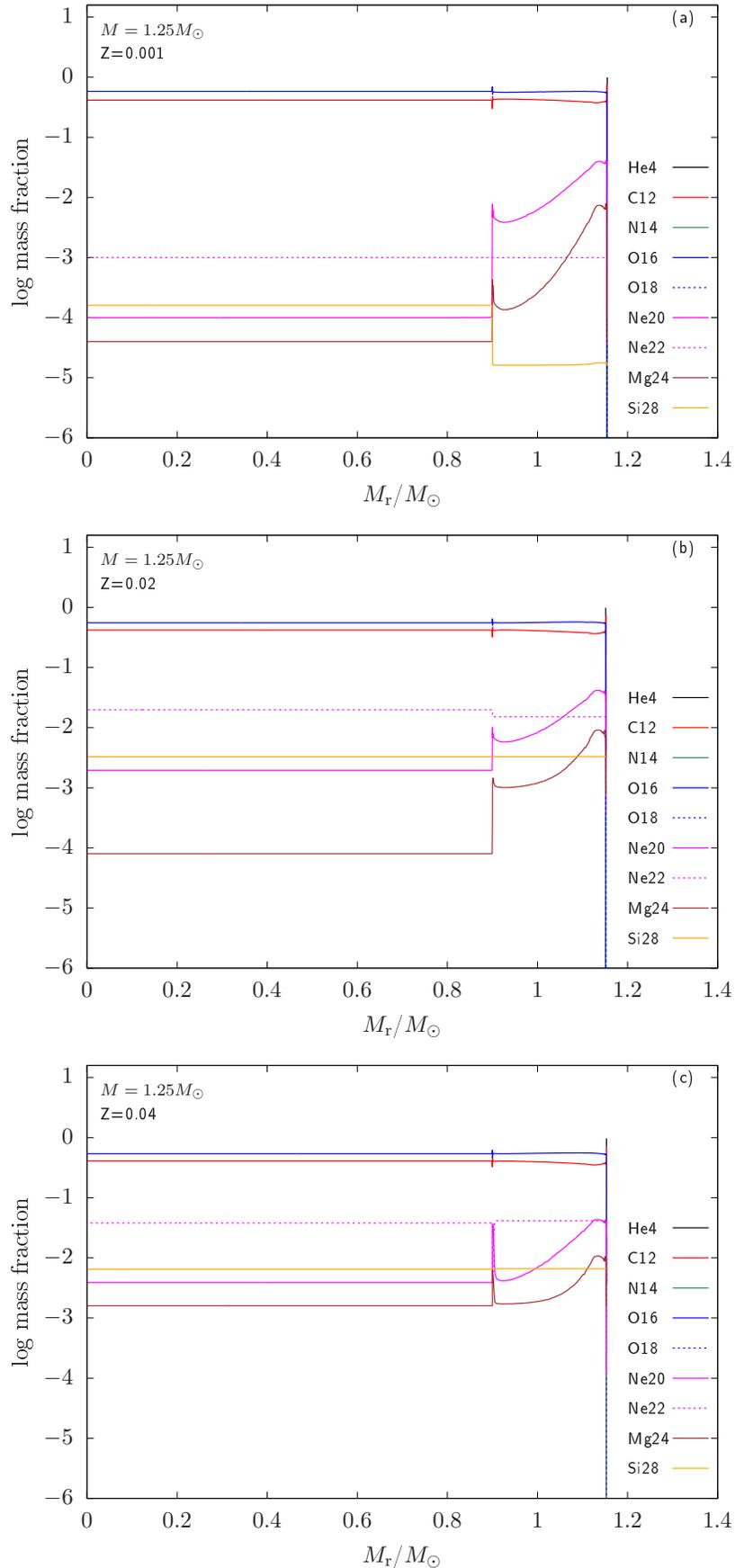,angle=0,width=12.2cm}
 \caption{Elemental abundance distributions for UMCOWDs resulted from $1.25{M}_\odot$ remnants with different initial metallicities.}
  \end{center}
\end{figure}

\subsection{Other uncertainties}

So far, several WD cooling mechanisms have been proposed to explain the extremely long cooling delays of the Q branch WDs (e.g. Bauer et al. 2020; Camisassa et al. 2021; Blouin, Daligault \& Saumon 2021), however, none of these mechanisms can account for over $6$-$8$ gigayears of cooling delay (e.g. Cheng, Cummings \& M\'enard 2019). One reason is that the WD cooling rate is mainly determined by its internal composition. Studies on the pulsation properties of WDs found that there exist markedly observable properties for the WDs with different internal or atmosphere compositions (e.g. De Ger\'onimo et al. 2019; C\'orsico et al. 2021; Chidester et al. 2021; Althaus et al. 2021). The difference in pulsation properties may be revealed through observations conducted with ongoing space missions and detailed asteroseismic analysis. Additionally, our work implies that the UMWDs in Q branch may have more complicated origins (i.e. evolving from single stars or from double WD mergers), making it more difficult to estimate the birthrate of the Q branch WDs and the double WD merger rate (e.g. Cheng, Cummings \& M\'enard 2019; Cheng et al. 2020; Camisassa et al. 2021). Therefore, further efforts on the WD cooling models, observations and simulations of asteroseismology are still needed.

\section{Summary}

By employing the stellar evolution code MESA, we constructed remnant models involving merger of a more massive CO WD with a He WD, and investigated their post-merger evolutions. We found that the evolution of the remnants is similar to R CrB stars but with higher luminosity. During the evolution, the inwardly extending convective region in the envelope can penetrate the He burning layer when the core mass increases above $1.15{M}_\odot$, leading to the decreases of the He abundance of the envelope in the subsequent evolution. Off-centre carbon ignition can be triggered when the core mass reaches about $1.2{M}_\odot$, and the ignition-mass of the core shows less relation to the initial CO core mass, metallicity of the progenitor, rotation and wind mass-loss rate of the remnant. The results of the current work indicated that the UMCOWD with mass no more than $1.2{M}_\odot$ can be produced from CO WD + He WD merger, which can partly explain the H-depletion UMWDs with long cooling delays existing in the Q branch.

\section*{Acknowledgments}
We thank the anonymous referee for his/her very helpful suggestions on the manuscript. We thank Even Bauer, Detlev Koester and Xianfei Zhang for helpful discussion. CYW is supported by the National Natural Science Foundation of China (NSFC grants 12003013). XFW is supported by the National Natural Science Foundation of China (NSFC grants 12033003, 11633002), the Scholar Program of Beijing Academy of Science and Technology (DZ:BS202002), and the tencent XPLORER prize.

\section*{DATA AVAILABILITY}
The source files for our mesa models along with the output files are publicly available on Zenodo at https://doi.org/10.5281/zenodo.5900481

\appendix
\section{Results of post-merger evolutions}

In this appendix, we display the major results of the post-merger evolutions which are listed in Table.\,A1-A4.

\begin{center}
\begin{table}
%\centering
\textbf{\caption{Information for the properties of various remnants. ${\rm Z}$=initial metallicity of the progenitors; ${\rm {{\eta}_{\rm R}}}$=Reimers' wind factor of the remnants during the giant phase applied in our simulations; ${{\Omega}/{\Omega}_{\rm crit}}$=initial rotation rate of the remnants; $\Delta{t}$=lifetime of the remnants showing as R CrB phase in unit of years; $\overline{\dot{M}}$=average mass-loss rates of the remnants during R CrB phase in unit of ${M}_\odot/{\rm yr}$; ${M}_{\rm F}$=final masses of WDs in unit of solar mass (for the remnants that experience off-centre carbon ignition, the corresponding value represents the core mass when carbon ignition occurs); ${M}_{\rm ig}$=total mass of the remnants when carbon ignition occurs in unit of solar mass; the last eight columns represent the average elemental abundance of the WDs (envelope removed). Items in black, blue and red typefaces represent the final outcomes of the merger remnants are CO WDs with masses less than $1.05{M}_\odot$, UMCOWDs, and experiencing off-central carbon ignitions, respectively.}}
\begin{tabular}{|c|c|c|c|c|c|c|c|c|c|c|c|c|c|c|c|}     % 8 columns
\hline
  ${\rm Z}$ & Model & ${\rm {{\eta}_{\rm R}}}$ & ${{\Omega}/{\Omega}_{\rm crit}}$ & $\Delta{t}$ & $\overline{\dot{M}}$ & ${M}_{\rm F}$ & ${M}_{\rm ig}$ & $^{12}{\rm C}$ & $^{14}{\rm N}$ & $^{16}{\rm O}$ & $^{18}{\rm O}$ & $^{20}{\rm Ne}$ & $^{22}{\rm Ne}$ & $^{24}{\rm Mg}$ & $^{28}{\rm Si}$\\
  \hline
  \hline
  0.02 & 0.7+0.35 & 0.5 & 0.0 & $1.31\times{10}^5$ & $8.68\times{10}^{-7}$ & 0.94 & - - & 43.14 & 0.0 & 54.44 & 0.0 & 0.21 & 1.86 & 0.03 & 0.33\\
       & 0.7+0.40 &     &  & $1.46\times{10}^{5}$ & $9.03\times{10}^{-7}$ & 0.98 & - - & 43.07 & 0.0 & 54.51 & 0.0 & 0.21 & 1.85 & 0.03 & 0.33\\
       & 0.7+0.45 &     &  & $1.58\times{10}^{5}$ & $9.34\times{10}^{-7}$ & 1.01 & - - & 43.01 & 0.0 & 54.57 & 0.0 & 0.22 & 1.84 & 0.03 & 0.33\\
       & 0.8+0.25 &     &  & $7.10\times{10}^{4}$ & $1.15\times{10}^{-6}$ & 0.97 & - - & 42.42 & 0.0 & 55.11 & 0.0 & 0.22 & 1.90 & 0.02 & 0.33\\
       & 0.8+0.30 &     &  & $8.31\times{10}^{4}$ & $1.20\times{10}^{-6}$ & 1.00 & - - & 42.39 & 0.0 & 55.14 & 0.0 & 0.23 & 1.89 & 0.02 & 0.33\\
       & 0.8+0.35 &     &  & $9.54\times{10}^{4}$ & $1.22\times{10}^{-6}$ & 1.04 & - - & 42.33 & 0.0 & 55.19 & 0.0 & 0.25 & 1.88 & 0.03 & 0.33\\
       & {\color{blue}0.8+0.40} &     &  & {\color{blue}$1.08\times{10}^{5}$} & {\color{blue}$1.23\times{10}^{-6}$} & {\color{blue}1.08} & - - & {\color{blue}42.22} & {\color{blue}0.0} & {\color{blue}55.26} & {\color{blue}0.0} & {\color{blue}0.29} & {\color{blue}1.86} & {\color{blue}0.033} & {\color{blue}0.33}\\
       & {\color{blue}0.8+0.45} &     &  & {\color{blue}$1.16\times{10}^{5}$} & {\color{blue}$1.23\times{10}^{-6}$} & {\color{blue}1.11} & - - & {\color{blue}42.11} & {\color{blue}0.0} & {\color{blue}55.33} & {\color{blue}0.0} & {\color{blue}0.34} & {\color{blue}1.85} & {\color{blue}0.04} & {\color{blue}0.33}\\
       & 0.9+0.20 &     &  & $4.86\times{4}$ & $1.27\times{10}^{-6}$ & 1.04 & - - & 41.97 & 0.0 & 55.49 & 0.0 & 0.26 & 1.92 & 0.022 & 0.33\\
       & {\color{blue}0.9+0.25} &     &  & {\color{blue}$6.0\times{10}^{4}$} & {\color{blue}$1.27\times{10}^{-6}$} & {\color{blue}1.08} & - - & {\color{blue}41.87} & {\color{blue}0.0} & {\color{blue}55.56} & {\color{blue}0.0} & {\color{blue}0.31} & {\color{blue}1.91} & {\color{blue}0.03} & {\color{blue}0.33}\\
       & {\color{blue}0.9+0.30} &     &  & {\color{blue}$7.06\times{10}^{4}$} & {\color{blue}$1.27\times{10}^{-6}$} & {\color{blue}1.12} & - - & {\color{blue}41.73} & {\color{blue}0.0} & {\color{blue}55.63} & {\color{blue}0.0} & {\color{blue}0.38} & {\color{blue}1.89} & {\color{blue}0.04} & {\color{blue}0.33}\\
       & {\color{blue}0.9+0.35} &     &  & {\color{blue}$7.92\times{10}^{4}$} & {\color{blue}$1.27\times{10}^{-6}$} & {\color{blue}1.15} & - - & {\color{blue}41.55} & {\color{blue}0.0} & {\color{blue}55.67} & {\color{blue}0.0} & {\color{blue}0.51} & {\color{blue}1.88} & {\color{blue}0.07} & {\color{blue}0.33}\\
       & {\color{blue}0.9+0.40} &     &  & {\color{blue}$8.76\times{10}^{4}$} & {\color{blue}$1.28\times{10}^{-6}$} & {\color{blue}1.19} & - - & {\color{blue}41.26} & {\color{blue}0.0} & {\color{blue}55.61} & {\color{blue}0.0} & {\color{blue}0.77} & {\color{blue}1.87} & {\color{blue}0.17} & {\color{blue}0.33}\\
       & {\color{red}0.9+0.45} &     &  & {\color{red}$8.83\times{10}^{4}$} & {\color{red}$1.38\times{10}^{-6}$} & {\color{red}1.20} & {\color{red}1.24} & - -  & - - & - -    & - - & - -   & - -   & - -    & - - \\
       & {\color{blue}1.0+0.20} &     &  & {\color{blue}$4.17\times{10}^{4}$} & {\color{blue}$1.16\times{10}^{-6}$} & {\color{blue}1.15} & - - & {\color{blue}41.45} & {\color{blue}0.0} & {\color{blue}55.67} & {\color{blue}0.0} & {\color{blue}0.55} & {\color{blue}1.92} & {\color{blue}0.08} & {\color{blue}0.33}\\
       & {\color{blue}1.0+0.25} &     &  & {\color{blue}$4.98\times{10}^{4}$} & {\color{blue}$1.22\times{10}^{-6}$} & {\color{blue}1.19} & - - & {\color{blue}41.16} & {\color{blue}0.0} & {\color{blue}55.58} & {\color{blue}0.0} & {\color{blue}0.83} & {\color{blue}1.91} & {\color{blue}0.19} & {\color{blue}0.33}\\
       & {\color{red}1.0+0.30} &     &  & {\color{red}$5.0\times{10}^{4}$} & {\color{red}$1.42\times{10}^{-6}$} & {\color{red}1.20} & {\color{red}1.23} & - -  & - - & - -    & - - & - -   & - -   & - -    & - - \\
       & {\color{red}1.0+0.35} &     &  & {\color{red}$4.8\times{10}^{4}$} & {\color{red}$1.68\times{10}^{-6}$} & {\color{red}1.20} & {\color{red}1.28} & - -  & - - & - -    & - - & - -   & - -   & - -    & - - \\
  \hline
  \hline
\end{tabular}
\end{table}
\end{center}

\begin{center}
\begin{table}
%\centering
\textbf{\caption{Similar to Table.\,A1, but for various wind mass-loss prescriptions (${\eta}_{\rm R}$ represents the scaling factor used in Reimers' wind prescription, whereas ${\eta}_{\rm B}$ represents the scaling factor used in Bl\"ocker's wind prescription).}}
\begin{tabular}{|c|c|c|c|c|c|c|c|c|c|c|c|c|c|c|c|}     % 8 columns
\hline
  ${\rm Z}$ & Model &  & ${{\Omega}/{\Omega}_{\rm crit}}$ & $\Delta{t}$ & $\overline{\dot{M}}$ & ${M}_{\rm F}$ & ${M}_{\rm ig}$ & $^{12}{\rm C}$ & $^{14}{\rm N}$ & $^{16}{\rm O}$ & $^{18}{\rm O}$ & $^{20}{\rm Ne}$ & $^{22}{\rm Ne}$ & $^{24}{\rm Mg}$ & $^{28}{\rm Si}$\\
  \hline
  \hline
    0.02 & 0.7+0.35 & ${\eta}_{\rm R}=0.05$ & 0.0 & $1.69\times{10}^{5}$ & $9.70\times{10}^{-8}$ & 1.04 & - - & 42.98 & 0.0 & 54.59 & 0.0 & 0.24 & 1.83 & 0.03 & 0.33\\
       & {\color{blue}0.7+0.40} &  &  & {\color{blue}$1.85\times{10}^{5}$} & {\color{blue}$9.90\times{10}^{-8}$} & {\color{blue}1.09} & - - & {\color{blue}42.79} & {\color{blue}0.0} & {\color{blue}54.73} & {\color{blue}0.0} & {\color{blue}0.30} & {\color{blue}1.82} & {\color{blue}0.04} & {\color{blue}0.33}\\
       & {\color{blue}0.7+0.45} &  &  & {\color{blue}$1.97\times{10}^{5}$} & {\color{blue}$1.00\times{10}^{-7}$} & {\color{blue}1.14} & - - & {\color{blue}42.54} & {\color{blue}0.0} & {\color{blue}54.85} & {\color{blue}0.0} & {\color{blue}0.41} & {\color{blue}1.80} & {\color{blue}0.06} & {\color{blue}0.33}\\
       & 0.8+0.25 &  &  & $9.68\times{10}^{4}$ & $1.20\times{10}^{-7}$ & 1.04 & - - & 42.34 & 0.0 & 55.18 & 0.0 & 0.25 & 1.88 & 0.03 & 0.33\\
       & {\color{blue}0.8+0.30} &  &  & {\color{blue}$1.11\times{10}^{5}$} & {\color{blue}$1.22\times{10}^{-7}$} & {\color{blue}1.09} & - - & {\color{blue}42.20} & {\color{blue}0.0} & {\color{blue}55.27} & {\color{blue}0.0} & {\color{blue}0.31} & {\color{blue}1.86} & {\color{blue}0.04} & {\color{blue}0.33}\\
       & {\color{blue}0.8+0.35} &  &  & {\color{blue}$1.24\times{10}^{5}$} & {\color{blue}$1.23\times{10}^{-7}$} & {\color{blue}1.14} & - - & {\color{blue}41.99} & {\color{blue}0.0} & {\color{blue}55.36} & {\color{blue}0.0} & {\color{blue}0.42} & {\color{blue}1.86} & {\color{blue}0.05} & {\color{blue}0.33}\\
       & {\color{blue}0.8+0.40} &  &  & {\color{blue}$1.36\times{10}^{5}$} & {\color{blue}$1.24\times{10}^{-7}$} & {\color{blue}1.19} & - - & {\color{blue}41.58} & {\color{blue}0.0} & {\color{blue}55.33} & {\color{blue}0.0} & {\color{blue}0.76} & {\color{blue}1.83} & {\color{blue}0.17} & {\color{blue}0.33}\\
       & {\color{red}0.8+0.45} &  &  & {\color{red}$1.38\times{10}^{5}$} & {\color{red}$1.29\times{10}^{-7}$} & {\color{red}1.21} & {\color{red}1.23} & - - & - - & - - & - - & - - & - - & - - & - -\\
       & {\color{blue}0.9+0.20} &  &  & {\color{blue}$6.43\times{10}^{4}$} & {\color{blue}$1.26\times{10}^{-7}$} & {\color{blue}1.09} & - - & {\color{blue}41.83} & {\color{blue}0.0} & {\color{blue}55.58} & {\color{blue}0.0} & {\color{blue}0.33} & {\color{blue}1.90} & {\color{blue}0.03} & {\color{blue}0.33}\\
       & {\color{blue}0.9+0.25} &  &  & {\color{blue}$7.75\times{10}^{4}$} & {\color{blue}$1.27\times{10}^{-7}$} & {\color{blue}1.14} & - - & {\color{blue}41.60} & {\color{blue}0.0} & {\color{blue}55.65} & {\color{blue}0.0} & {\color{blue}0.47} & {\color{blue}1.88} & {\color{blue}0.06} & {\color{blue}0.33}\\
       & {\color{blue}0.9+0.30} &  &  & {\color{blue}$8.84\times{10}^{4}$} & {\color{blue}$1.27\times{10}^{-7}$} & {\color{blue}1.20} & - - & {\color{blue}41.24} & {\color{blue}0.0} & {\color{blue}55.58} & {\color{blue}0.0} & {\color{blue}0.81} & {\color{blue}1.87} & {\color{blue}0.18} & {\color{blue}0.33}\\
       & {\color{red}0.9+0.35} &  &  & {\color{red}$8.82\times{10}^{4}$} & {\color{red}$1.40\times{10}^{-7}$} & {\color{red}1.20} & {\color{red}1.24} & - - & - - & - - & - - & - - & - - & - - & - -\\
       & {\color{red}0.9+0.40} &  &  & {\color{red}$8.64\times{10}^{4}$} & {\color{red}$1.55\times{10}^{-7}$} & {\color{red}1.20} & {\color{red}1.29} & - - & - - & - - & - - & - - & - - & - - & - -\\
       & {\color{red}0.9+0.45} &  &  & {\color{red}$8.61\times{10}^{4}$} & {\color{red}$1.69\times{10}^{-7}$} & {\color{red}1.20} & {\color{red}1.34} & - - & - - & - - & - - & - - & - - & - - & - -\\
       & {\color{red}1.0+0.25} &  &  & {\color{red}$4.63\times{10}^{4}$} & {\color{red}$1.49\times{10}^{-7}$} & {\color{red}1.20} & {\color{red}1.25} & - - & - - & - - & - - & - - & - - & - - & - -\\
  \hline
  0.02 & 0.7+0.35 & ${\eta}_{\rm R}=5.0$ & 0.0 & $3.76\times{10}^{4}$ & $7.47\times{10}^{-6}$ & 0.77 & - - & 42.68 & 0.0 & 54.84 & 0.0 & 0.20 & 1.94 & 0.01 & 0.33\\
       & 0.7+0.40 &  &  & $4.46\times{10}^{4}$ & $7.28\times{10}^{-6}$ & 0.78 & - - & 42.72 & 0.0 & 54.81 & 0.0 & 0.20 & 1.94 & 0.02 & 0.33\\
       & 0.7+0.45 &  &  & $5.07\times{10}^{4}$ & $7.20\times{10}^{-6}$ & 0.79 & - - & 42.76 & 0.0 & 54.77 & 0.0 & 0.20 & 1.93 & 0.02 & 0.33\\
       & 0.8+0.25 &  &  & $1.91\times{10}^{4}$ & $1.05\times{10}^{-5}$ & 0.85 & - - & 42.25 & 0.0 & 55.25 & 0.0 & 0.20 & 1.96 & 0.01 & 0.33\\
       & 0.8+0.30 &  &  & $2.27\times{10}^{4}$ & $1.07\times{10}^{-5}$ & 0.86 & - - & 42.27 & 0.0 & 55.24 & 0.0 & 0.20 & 1.95 & 0.01 & 0.33\\
       & 0.8+0.35 &  &  & $2.63\times{10}^{4}$ & $1.09\times{10}^{-5}$ & 0.87 & - - & 42.28 & 0.0 & 55.23 & 0.0 & 0.20 & 1.95 & 0.01 & 0.33\\
       & 0.8+0.40 &  &  & $3.04\times{10}^{4}$ & $1.10\times{10}^{-5}$ & 0.88 & - - & 42.30 & 0.0 & 55.22 & 0.0 & 0.20 & 1.94 & 0.01 & 0.33\\
       & 0.8+0.45 &  &  & $3.34\times{10}^{4}$ & $1.11\times{10}^{-5}$ & 0.88 & - - & 42.31 & 0.0 & 55.20 & 0.0 & 0.20 & 1.94 & 0.01 & 0.33\\
       & 0.9+0.20 &  &  & $1.28\times{10}^{4}$ & $1.27\times{10}^{-5}$ & 0.94 & - - & 42.03 & 0.0 & 55.45 & 0.0 & 0.21 & 1.96 & 0.01 & 0.33\\
       & 0.9+0.25 &  &  & $1.59\times{10}^{4}$ & $1.29\times{10}^{-5}$ & 0.95 & - - & 42.04 & 0.0 & 55.45 & 0.0 & 0.21 & 1.96 & 0.01 & 0.33\\
       & 0.9+0.30 &  &  & $1.90\times{10}^{4}$ & $1.31\times{10}^{-5}$ & 0.96 & - - & 42.03 & 0.0 & 55.45 & 0.0 & 0.22 & 1.95 & 0.01 & 0.33\\
       & 0.9+0.35 &  &  & $2.18\times{10}^{4}$ & $1.32\times{10}^{-5}$ & 0.96 & - - & 42.03 & 0.0 & 55.46 & 0.0 & 0.22 & 1.95 & 0.01 & 0.33\\
       & 0.9+0.40 &  &  & $2.49\times{10}^{4}$ & $1.33\times{10}^{-5}$ & 0.97 & - - & 42.03 & 0.0 & 55.46 & 0.0 & 0.22 & 1.95 & 0.01 & 0.33\\
       & 0.9+0.45 &  &  & $2.83\times{10}^{4}$ & $1.33\times{10}^{-5}$ & 0.98 & - - & 42.02 & 0.0 & 55.46 & 0.0 & 0.23 & 1.94 & 0.02 & 0.33\\
       & {\color{blue}1.0+0.25} &  &  & {\color{blue}$1.64\times{10}^{4}$} & {\color{blue}$1.18\times{10}^{-5}$} & {\color{blue}1.06} & - - & {\color{blue}41.84} & {\color{blue}0.0} & {\color{blue}55.57} & {\color{blue}0.0} & {\color{blue}0.29} & {\color{blue}1.96} & {\color{blue}0.02} & {\color{blue}0.33}\\
       & {\color{blue}1.0+0.30} &  &  & {\color{blue}$1.91\times{10}^{4}$} & {\color{blue}$1.23\times{10}^{-5}$} & {\color{blue}1.07} & - - & {\color{blue}41.81} & {\color{blue}0.0} & {\color{blue}55.59} & {\color{blue}0.0} & {\color{blue}0.30} & {\color{blue}1.95} & {\color{blue}0.03} & {\color{blue}0.33}\\
       & {\color{blue}1.0+0.45} &  &  & {\color{blue}$2.75\times{10}^{4}$} & {\color{blue}$1.28\times{10}^{-5}$} & {\color{blue}1.10} & - - & {\color{blue}41.68} & {\color{blue}0.0} & {\color{blue}55.65} & {\color{blue}0.0} & {\color{blue}0.37} & {\color{blue}1.94} & {\color{blue}0.04} & {\color{blue}0.33}\\
  \hline
  0.02 & 0.7+0.35 & ${\eta}_{\rm B}=0.02$ & 0.0 & $3.54\times{10}^{3}$ & $9.53\times{10}^{-5}$ & 0.72 & - - & 42.17 & 0.0 & 55.33 & 0.0 & 0.20 & 1.97 & 0.01 & 0.33\\
       & 0.7+0.40 &  &  & $4.59\times{10}^{3}$ & $8.43\times{10}^{-5}$ & 0.72 & - - & 42.20 & 0.0 & 55.29 & 0.0 & 0.20 & 1.97 & 0.01 & 0.33\\
       & 0.7+0.45 &  &  & $3.56\times{10}^{3}$ & $7.36\times{10}^{-5}$ & 0.72 & - - & 42.23 & 0.0 & 55.26 & 0.0 & 0.20 & 1.97 & 0.01 & 0.33\\
       & 0.8+0.25 &  &  & $8.79\times{10}^{2}$ & $2.77\times{10}^{-4}$ & 0.81 & - - & 42.02 & 0.0 & 55.47 & 0.0 & 0.20 & 1.99 & 0.01 & 0.33\\
       & 0.8+0.30 &  &  & $1.02\times{10}^{3}$ & $2.87\times{10}^{-4}$ & 0.81 & - - & 42.03 & 0.0 & 55.46 & 0.0 & 0.20 & 1.98 & 0.01 & 0.33\\
       & 0.8+0.35 &  &  & $1.25\times{10}^{3}$ & $2.72\times{10}^{-4}$ & 0.81 & - - & 42.03 & 0.0 & 55.45 & 0.0 & 0.20 & 1.98 & 0.01 & 0.33\\
       & 0.8+0.40 &  &  & $1.50\times{10}^{3}$ & $2.67\times{10}^{-4}$ & 0.81 & - - & 42.04 & 0.0 & 55.45 & 0.0 & 0.20 & 1.98 & 0.01 & 0.33\\
       & 0.8+0.45 &  &  & $1.74\times{10}^{3}$ & $2.53\times{10}^{-4}$ & 0.81 & - - & 42.05 & 0.0 & 55.44 & 0.0 & 0.20 & 1.98 & 0.01 & 0.33\\
       & 0.9+0.20 &  &  & $9.60\times{10}^{2}$ & $2.06\times{10}^{-4}$ & 0.90 & - - & 42.00 & 0.0 & 55.50 & 0.0 & 0.20 & 1.98 & 0.01 & 0.33\\
       & 0.9+0.25 &  &  & $1.08\times{10}^{3}$ & $2.33\times{10}^{-4}$ & 0.90 & - - & 42.00 & 0.0 & 55.50 & 0.0 & 0.20 & 1.98 & 0.01 & 0.33\\
       & 0.9+0.30 &  &  & $1.20\times{10}^{3}$ & $2.52\times{10}^{-4}$ & 0.90 & - - & 42.00 & 0.0 & 55.50 & 0.0 & 0.20 & 1.98 & 0.01 & 0.33\\
       & 0.9+0.35 &  &  & $1.31\times{10}^{3}$ & $2.66\times{10}^{-4}$ & 0.90 & - - & 42.00 & 0.0 & 55.50 & 0.0 & 0.20 & 1.98 & 0.01 & 0.33\\
       & 0.9+0.40 &  &  & $6.27\times{10}^{2}$ & $6.34\times{10}^{-4}$ & 0.91 & - - & 42.00 & 0.0 & 55.50 & 0.0 & 0.20 & 1.98 & 0.01 & 0.33\\
       & 0.9+0.45 &  &  & $7.67\times{10}^{2}$ & $5.88\times{10}^{-4}$ & 0.91 & - - & 42.00 & 0.0 & 55.50 & 0.0 & 0.20 & 1.98 & 0.01 & 0.33\\
       & 1.0+0.25 &  &  & $4.36\times{10}^{2}$ & $5.73\times{10}^{-4}$ & 1.00 & - - & 42.00 & 0.0 & 55.50 & 0.0 & 0.20 & 1.98 & 0.01 & 0.33\\
       & 1.0+0.30 &  &  & $4.68\times{10}^{2}$ & $6.42\times{10}^{-4}$ & 1.00 & - - & 42.00 & 0.0 & 55.50 & 0.0 & 0.20 & 1.98 & 0.01 & 0.33\\
       & 1.0+0.35 &  &  & $5.40\times{10}^{2}$ & $6.52\times{10}^{-4}$ & 1.00 & - - & 42.00 & 0.0 & 55.50 & 0.0 & 0.20 & 1.98 & 0.01 & 0.33\\
       & 1.0+0.45 &  &  & $6.24\times{10}^{2}$ & $7.21\times{10}^{-4}$ & 1.00 & - - & 42.00 & 0.0 & 55.50 & 0.0 & 0.20 & 1.98 & 0.01 & 0.33\\
  \hline
    \hline
\end{tabular}
\end{table}
\end{center}

\begin{center}
\begin{table}
%\centering
\textbf{\caption{Similar to Table.\,A1, but for considering rotation.}}
\begin{tabular}{|c|c|c|c|c|c|c|c|c|c|c|c|c|c|c|c|}     % 8 columns
\hline
  ${\rm Z}$ & Model & ${\rm {{\eta}_{\rm R}}}$ & ${{\Omega}/{\Omega}_{\rm crit}}$ & $\Delta{t}$ & $\overline{\dot{M}}$ & ${M}_{\rm F}$ & ${M}_{\rm ig}$ & $^{12}{\rm C}$ & $^{14}{\rm N}$ & $^{16}{\rm O}$ & $^{18}{\rm O}$ & $^{20}{\rm Ne}$ & $^{22}{\rm Ne}$ & $^{24}{\rm Mg}$ & $^{28}{\rm Si}$\\
  \hline
  \hline
  0.02 & 0.8+0.30 & 0.5 & 0.4 & $7.97\times{10}^{4}$ & $1.41\times{10}^{-6}$ & 0.99 & - - & 41.65 & 0.0 & 55.86 & 0.0 & 0.25 & 1.89 & 0.02 & 0.33\\
       & 0.8+0.35 &  &  & $8.93\times{10}^{4}$ & $1.47\times{10}^{-6}$ & 1.02 & - - & 41.37 & 0.0 & 56.10 & 0.0 & 0.28 & 1.88 & 0.03 & 0.33\\
       & {\color{blue}0.8+0.40} &  &  & {\color{blue}$1.04\times{10}^{5}$} & {\color{blue}$1.46\times{10}^{-6}$} & {\color{blue}1.06} & - - & {\color{blue}41.24} & {\color{blue}0.0} & {\color{blue}56.21} & {\color{blue}0.0} & {\color{blue}0.31} & {\color{blue}1.87} & {\color{blue}0.03} & {\color{blue}0.33}\\
       & {\color{blue}0.8+0.45} &  &  & {\color{blue}$1.12\times{10}^{5}$} & {\color{blue}$1.46\times{10}^{-6}$} & {\color{blue}1.09} & - - & {\color{blue}41.13} & {\color{blue}0.0} & {\color{blue}56.29} & {\color{blue}0.0} & {\color{blue}0.36} & {\color{blue}1.86} & {\color{blue}0.04} & {\color{blue}0.33}\\
       & 0.9+0.20 &  &  & $4.54\times{10}^{4}$ & $1.57\times{10}^{-6}$ & 1.03 & - - & 41.56 & 0.0 & 55.88 & 0.0 & 0.28 & 1.92 & 0.02 & 0.33\\
       & {\color{blue}0.9+0.30} &  &  & {\color{blue}$6.65\times{10}^{4}$} & {\color{blue}$1.60\times{10}^{-6}$} & {\color{blue}1.10} & - - & {\color{blue}41.17} & {\color{blue}0.0} & {\color{blue}56.16} & {\color{blue}0.0} & {\color{blue}0.40} & {\color{blue}1.90} & {\color{blue}0.04} & {\color{blue}0.33}\\
       & {\color{blue}0.9+0.35} &  &  & {\color{blue}$7.54\times{10}^{4}$} & {\color{blue}$1.58\times{10}^{-6}$} & {\color{blue}1.13} & - - & {\color{blue}40.97} & {\color{blue}0.0} & {\color{blue}56.24} & {\color{blue}0.0} & {\color{blue}0.52} & {\color{blue}1.89} & {\color{blue}0.07} & {\color{blue}0.33}\\
       & {\color{blue}0.9+0.40} &  &  & {\color{blue}$8.40\times{10}^{4}$} & {\color{blue}$1.57\times{10}^{-6}$} & {\color{blue}1.17} & - - & {\color{blue}40.78} & {\color{blue}0.0} & {\color{blue}56.21} & {\color{blue}0.0} & {\color{blue}0.69} & {\color{blue}1.87} & {\color{blue}0.12} & {\color{blue}0.33}\\
       & {\color{red}0.9+0.45} &  &  & {\color{red}$9.19\times{10}^{4}$} & {\color{red}$1.57\times{10}^{-6}$} & {\color{red}1.21} & {\color{red}1.21} & - - & - - & - - & - - & - - & - - & - - & - - \\
       & {\color{blue}1.0+0.25} &  &  & {\color{blue}$4.87\times{10}^{4}$} & {\color{blue}$1.36\times{10}^{-6}$} & {\color{blue}1.19} & - - & {\color{blue}44.16} & {\color{blue}0.0} & {\color{blue}59.92} & {\color{blue}0.0} & {\color{blue}0.84} & {\color{blue}2.06} & {\color{blue}0.18} & {\color{blue}0.33}\\
       & {\color{blue}1.0+0.30} &  &  & {\color{blue}$5.48\times{10}^{4}$} & {\color{blue}$1.55\times{10}^{-6}$} & {\color{blue}1.22} & - - & {\color{blue}37.82} & {\color{blue}0.0} & {\color{blue}54.16} & {\color{blue}0.0} & {\color{blue}4.49} & {\color{blue}1.90} & {\color{blue}1.19} & {\color{blue}0.44}\\
  \hline
  \hline
\end{tabular}
\end{table}
\end{center}

\begin{center}
\begin{table}
%\centering
\textbf{\caption{Similar to Table.\,A1, but for various metallicities}}
\begin{tabular}{|c|c|c|c|c|c|c|c|c|c|c|c|c|c|c|c|}     % 8 columns
\hline
  ${\rm Z}$ & Model & ${\rm {{\eta}_{\rm R}}}$ & ${{\Omega}/{\Omega}_{\rm crit}}$ & $\Delta{t}$ & $\overline{\dot{M}}$ & ${M}_{\rm F}$ & ${M}_{\rm ig}$ & $^{12}{\rm C}$ & $^{14}{\rm N}$ & $^{16}{\rm O}$ & $^{18}{\rm O}$ & $^{20}{\rm Ne}$ & $^{22}{\rm Ne}$ & $^{24}{\rm Mg}$ & $^{28}{\rm Si}$\\
  \hline
  \hline
  0.001& 0.7+0.35 & 0.5 & 0.0 & $1.37\times{10}^{5}$ & $7.33\times{10}^{-7}$ & 0.95 & - - & 43.08 & 0.0 & 56.78 & 0.0 & 0.02 & 0.1 & 0.00 & 0.01\\
       & 0.7+0.40 &  &  & $1.53\times{10}^{5}$ & $7.59\times{10}^{-7}$ & 0.99 & - - & 43.07 & 0.0 & 56.78 & 0.0 & 0.03 & 0.10  & 0.00 & 0.01\\
       & 0.7+0.45 &  &  & $1.65\times{10}^{5}$ & $7.85\times{10}^{-7}$ & 1.03 & - - & 43.03 & 0.0 & 56.80 & 0.0 & 0.05 & 0.10  & 0.00 & 0.01\\
       & 0.8+0.25 &  &  & $7.45\times{10}^{4}$ & $1.03\times{10}^{-6}$ & 0.98 & - - & 42.31 & 0.0 & 57.54 & 0.0 & 0.04 & 0.10  & 0.00 & 0.01\\
       & 0.8+0.30 &  &  & $8.67\times{10}^{4}$ & $1.06\times{10}^{-6}$ & 1.01 & - - & 42.31 & 0.0 & 57.53 & 0.0 & 0.05 & 0.10  & 0.00 & 0.01\\
       & {\color{blue}0.8+0.35} &  &  & {\color{blue}$9.74\times{10}^{4}$} & {\color{blue}$1.10\times{10}^{-6}$} & {\color{blue}1.05} & - - & {\color{blue}42.27} & {\color{blue}0.0} & {\color{blue}57.54} & {\color{blue}0.0} & {\color{blue}0.07} & {\color{blue}0.10}  & {\color{blue}0.01} & {\color{blue}0.01}\\
       & {\color{blue}0.8+0.40} &  &  & {\color{blue}$1.07\times{10}^{5}$} & {\color{blue}$1.09\times{10}^{-6}$} & {\color{blue}1.09} & - - & {\color{blue}42.20} & {\color{blue}0.0} & {\color{blue}57.57} & {\color{blue}0.0} & {\color{blue}0.12} & {\color{blue}0.10}  & {\color{blue}0.01} & {\color{blue}0.01}\\
       & {\color{blue}0.8+0.45} &  &  & {\color{blue}$1.16\times{10}^{5}$} & {\color{blue}$1.09\times{10}^{-6}$} & {\color{blue}1.13} & - - & {\color{blue}42.06} & {\color{blue}0.0} & {\color{blue}57.60} & {\color{blue}0.0} & {\color{blue}0.20} & {\color{blue}0.10}  & {\color{blue}0.02} & {\color{blue}0.01}\\
       & 0.9+0.20 &  &  & $4.65\times{10}^{4}$ & $1.33\times{10}^{-6}$ & 1.04 & - - & 41.81 & 0.0 & 57.99 & 0.0 & 0.08 & 0.10  & 0.01 & 0.01\\
       & {\color{blue}0.9+0.25} &  &  & {\color{blue}$5.59\times{10}^{4}$} & {\color{blue}$1.31\times{10}^{-6}$} & {\color{blue}1.08} & - - & {\color{blue}41.77} & {\color{blue}0.0} & {\color{blue}57.99} & {\color{blue}0.0} & {\color{blue}0.12} & {\color{blue}0.10}  & {\color{blue}0.01} & {\color{blue}0.01}\\
       & {\color{blue}0.9+0.30} &  &  & {\color{blue}$6.58\times{10}^{4}$} & {\color{blue}$1.30\times{10}^{-6}$} & {\color{blue}1.12} & - - & {\color{blue}41.66} & {\color{blue}0.0} & {\color{blue}58.01} & {\color{blue}0.0} & {\color{blue}0.20} & {\color{blue}0.10}  & {\color{blue}0.02} & {\color{blue}0.01}\\
       & {\color{blue}0.9+0.35} &  &  & {\color{blue}$7.35\times{10}^{4}$} & {\color{blue}$1.29\times{10}^{-6}$} & {\color{blue}1.16} & - - & {\color{blue}41.41} & {\color{blue}0.0} & {\color{blue}58.01} & {\color{blue}0.0} & {\color{blue}0.32} & {\color{blue}0.10}  & {\color{blue}0.40} & {\color{blue}0.01}\\
       & {\color{blue}0.9+0.40} &  &  & {\color{blue}$8.33\times{10}^{4}$} & {\color{blue}$1.26\times{10}^{-6}$} & {\color{blue}1.20} & - - & {\color{blue}41.23} & {\color{blue}0.0} & {\color{blue}57.90} & {\color{blue}0.0} & {\color{blue}0.62} & {\color{blue}0.10}  & {\color{blue}0.14} & {\color{blue}0.01}\\
       & {\color{red}0.9+0.45} &  &  & {\color{red}$9.00\times{10}^{4}$} & {\color{red}$1.29\times{10}^{-6}$} & {\color{red}1.21} & {\color{red}1.25} & - - & - - & - - & - - & - - & - - & - - & - - \\
       & {\color{blue}1.0+0.20} &  &  & {\color{blue}$3.54\times{10}^{4}$} & {\color{blue}$1.30\times{10}^{-6}$} & {\color{blue}1.16} & - - & {\color{blue}41.28} & {\color{blue}0.0} & {\color{blue}58.19} & {\color{blue}0.0} & {\color{blue}0.37} & {\color{blue}0.10}  & {\color{blue}0.06} & {\color{blue}0.01}\\
       & {\color{red}1.0+0.30} &  &  & {\color{red}$5.22\times{10}^{4}$} & {\color{red}$1.31\times{10}^{-6}$} & {\color{red}1.21} & {\color{red}1.23} & - - & - - & - - & - - & - - & - - & - - & - - \\
       & {\color{red}1.0+0.35} &  &  & {\color{red}$4.99\times{10}^{4}$} & {\color{red}$1.58\times{10}^{-6}$} & {\color{red}1.20} & {\color{red}1.28} & - - & - - & - - & - - & - - & - - & - - & - - \\
  \hline
  0.04 & 0.7+0.35 & 0.5 & 0.0 & $1.25\times{10}^{5}$ & $1.00\times{10}^{-6}$ & 0.93 & - - & 42.23 & 0.0 & 52.77 & 0.0 & 0.30 & 3.89 & 0.16 & 0.65\\
       & 0.7+0.40 &  &  & $1.39\times{10}^{5}$ & $1.04\times{10}^{-6}$ & 0.96 & - - & 42.12 & 0.0 & 52.87 & 0.0 & 0.30 & 3.89 & 0.16 & 0.65\\
       & 0.7+0.45 &  &  & $1.50\times{10}^{5}$ & $1.09\times{10}^{-6}$ & 0.99 & - - & 41.98 & 0.0 & 53.00 & 0.0 & 0.30 & 3.90 & 0.16 & 0.65\\
       & 0.8+0.25 &  &  & $6.88\times{10}^{4}$ & $1.19\times{10}^{-6}$ & 0.97 & - - & 41.00 & 0.0 & 54.05 & 0.0 & 0.35 & 3.78 & 0.16 & 0.65\\
       & 0.8+0.30 &  &  & $8.13\times{10}^{4}$ & $1.24\times{10}^{-6}$ & 1.01 & - - & 40.97 & 0.0 & 54.07 & 0.0 & 0.35 & 3.79 & 0.16 & 0.65\\
       & 0.8+0.35 &  &  & $9.08\times{10}^{4}$ & $1.28\times{10}^{-6}$ & 1.04 & - - & 40.92 & 0.0 & 54.11 & 0.0 & 0.36 & 3.80 & 0.16 & 0.65\\
       & {\color{blue}0.8+0.40} &  &  & {\color{blue}$1.01\times{10}^{5}$} & {\color{blue}$1.30\times{10}^{-6}$} & {\color{blue}1.07} & - - & {\color{blue}40.82}  & {\color{blue}0.0} & {\color{blue}54.16}  & {\color{blue}0.0} & {\color{blue}0.39} & {\color{blue}3.81} & {\color{blue}0.16} & {\color{blue}0.65}\\
       & {\color{blue}0.8+0.45} &  &  & {\color{blue}$1.10\times{10}^{5}$} & {\color{blue}$1.30\times{10}^{-6}$} & {\color{blue}1.11} & - - & {\color{blue}40.67}  & {\color{blue}0.0} & {\color{blue}54.23}  & {\color{blue}0.0} & {\color{blue}0.45} & {\color{blue}3.83} & {\color{blue}0.17} & {\color{blue}0.65}\\
       & 0.9+0.20 &  &  & $4.44\times{10}^{4}$ & $1.36\times{10}^{-6}$ & 1.04 & - - & 40.93 & 0.0 & 54.00 & 0.0 & 0.42 & 3.84 & 0.16 & 0.65\\
       & {\color{blue}0.9+0.30} &  &  & {\color{blue}$6.24\times{10}^{4}$} & {\color{blue}$1.38\times{10}^{-6}$} & {\color{blue}1.11} & - - & {\color{blue}40.67}  & {\color{blue}0.0} & {\color{blue}54.13}  & {\color{blue}0.0} & {\color{blue}0.52} & {\color{blue}3.86} & {\color{blue}0.18} & {\color{blue}0.65}\\
       & {\color{blue}0.9+0.35} &  &  & {\color{blue}$7.04\times{10}^{4}$} & {\color{blue}$1.39\times{10}^{-6}$} & {\color{blue}1.15} & - - & {\color{blue}40.45}  & {\color{blue}0.0} & {\color{blue}54.16}  & {\color{blue}0.0} & {\color{blue}0.66} & {\color{blue}3.87} & {\color{blue}0.21} & {\color{blue}0.65}\\
       & {\color{blue}0.9+0.40} &  &  & {\color{blue}$8.10\times{10}^{4}$} & {\color{blue}$1.35\times{10}^{-6}$} & {\color{blue}1.19} & - - & {\color{blue}40.12}  & {\color{blue}0.0} & {\color{blue}54.04}  & {\color{blue}0.0} & {\color{blue}0.97}  & {\color{blue}3.88} & {\color{blue}0.34} & {\color{blue}0.65}\\
       & {\color{red}0.9+0.45} &  &  & {\color{red}$8.56\times{10}^{4}$} & {\color{red}$1.38\times{10}^{-6}$} & {\color{red}1.20} & {\color{red}1.23} & - - & - - & - - & - - & - - & - - & - - & - - \\
       & {\color{blue}1.0+0.25} &  &  & {\color{blue}$4.33\times{10}^{4}$} & {\color{blue}$1.36\times{10}^{-6}$} & {\color{blue}1.19} & - - & {\color{blue}37.86}  & {\color{blue}0.0} & {\color{blue}56.22}  & {\color{blue}0.0} & {\color{blue}1.04}  & {\color{blue}3.87} & {\color{blue}0.36} & {\color{blue}0.65}\\
       & {\color{red}1.0+0.30} &  &  & {\color{red}$4.77\times{10}^{4}$} & {\color{red}$1.45\times{10}^{-6}$} & {\color{red}1.20} & {\color{red}1.23} & - - & - - & - - & - - & - - & - - & - - & - - \\
       & {\color{red}1.0+0.40} &  &  & {\color{red}$4.57\times{10}^{4}$} & {\color{red}$1.51\times{10}^{-6}$} & {\color{red}1.19} & {\color{red}1.33} & - - & - - & - - & - - & - - & - - & - - & - - \\
  \hline
  \hline
\end{tabular}
\end{table}
\end{center}

\label{lastpage}
\end{document}